\documentclass[11pt,a4paper]{article}
\usepackage[colorlinks=true, linkcolor=black!50!blue, urlcolor=blue, citecolor=blue, anchorcolor=blue]{hyperref}
\usepackage[font=small,labelfont=bf,margin=0mm,labelsep=period,tableposition=top]{caption}
\usepackage[a4paper,top=3cm,bottom=2.5cm,left=2.5cm,right=2.5cm,bindingoffset=0mm]{geometry}

\usepackage{graphicx,placeins}
\usepackage{float}
\usepackage{afterpage}
\usepackage{epsfig,cite}
\usepackage{amssymb}
\usepackage{amsmath}
\usepackage{dsfont}
\usepackage{multirow}
\usepackage{url}
\usepackage{xcolor,colortbl}
\usepackage{float}
\usepackage{afterpage}
\usepackage{url}
\usepackage{hyperref}
\usepackage{booktabs}
\usepackage{mathrsfs}
\usepackage{subcaption}

\usepackage{tikz}
\usepackage{tikz-3dplot}
\usepackage[compat=1.0.0]{tikz-feynman}

\usepackage{enumitem}
\usepackage{hyperref}
\usepackage{cite}

\usepackage{pifont}

\usetikzlibrary{shapes, arrows}
\usetikzlibrary{decorations.pathreplacing}
\usetikzlibrary{positioning, calc}
\tikzstyle{fitted} = [rectangle, minimum width=5cm, minimum height=1cm, text centered, draw=black, fill=red!30]
\tikzstyle{operations} = [rectangle, rounded corners, minimum width=2cm,text centered, draw=black, fill=red!30]
\tikzstyle{roundtext} = [rectangle, rounded corners, minimum width=2cm, minimum height=0.8cm, text centered, draw=black, fill=red!30]
\tikzstyle{n3py} = [rectangle, rounded corners, minimum width=3cm, minimum height=1cm, text centered, draw=black, fill=green!30]
\tikzstyle{myarrow} = [thick,->,>=stealth]
\tikzstyle{line} =[draw, -latex']
\tikzstyle{decision} = [diamond, draw, fill=red!20, text width=7.5em, text centered,  inner sep=0pt, minimum height=2em, aspect=4]
\tikzstyle{cloud} = [draw, ellipse,fill=green!20, minimum height=2em]
\tikzstyle{inout} = [rectangle, draw, fill=green!20, text width=9.5em, text centered, rounded corners, minimum height=2em, minimum width=10em]
\tikzstyle{block}=[rectangle, draw, fill=blue!20, text width=9.5em, 
                   text centered, rounded corners, minimum height=2em, 
                   minimum width=10em]

\definecolor{darkgreen}{rgb}{0.0, 0.5, 0.13}

\newcommand{\be}{\begin{equation}}
\newcommand{\ee}{\end{equation}}
\newcommand{\bea}{\begin{eqnarray}}
\newcommand{\eea}{\end{eqnarray}}
\newcommand{\bi}{\begin{itemize}}
\newcommand{\ei}{\end{itemize}}
\newcommand{\ben}{\begin{enumerate}}
\newcommand{\een}{\end{enumerate}}

\def\gsim{\mathrel{\rlap{\lower4pt\hbox{\hskip1pt$\sim$}}
    \raise1pt\hbox{$>$}}}         
\def\lsim{\mathrel{\rlap{\lower4pt\hbox{\hskip1pt$\sim$}}
    \raise1pt\hbox{$<$}}}         

\newcommand{\draft}[1]{}

\def\beq{\begin{equation}}
\def\eeq{\end{equation}}

\def \a{\alpha}

\def\lapprox{\lower .7ex\hbox{$\;\stackrel{\textstyle <}{\sim}\;$}}
\def\gapprox{\lower .7ex\hbox{$\;\stackrel{\textstyle >}{\sim}\;$}}

\def\d{{\rm d}}

\numberwithin{equation}{section}
\numberwithin{figure}{section}
\numberwithin{table}{section}

\usepackage{tabularx}
\newcolumntype{C}[1]{>{\centering\arraybackslash}p{#1}}

\usepackage{amsmath}
\usepackage{ gensymb }
\usepackage{amsfonts}
\usepackage{amssymb}
\usepackage{dsfont}
\usepackage{pifont}
\usepackage{booktabs}
\usepackage{graphicx}
\usepackage{epstopdf}
\usepackage{epsfig}
\usepackage{framed}
\usepackage{makeidx}
\usepackage{siunitx}
\usepackage[capitalise]{cleveref}
\usepackage{hyperref}
\usepackage{placeins}
\usepackage[font=small,labelfont=bf]{caption}

\RequirePackage{csquotes}
\usepackage{bbm}

\begin{document}
\newgeometry{top=1.5cm,bottom=1.5cm,left=1.5cm,right=1.5cm,bindingoffset=0mm}

\vspace{-2.0cm}
\begin{flushright}
TIF-UNIMI-2023-22\\
Edinburgh 2023/34\\
CERN-TH-2024-009\\
\end{flushright}
\vspace{0.3cm}

\begin{center}
  {\Large \bf Determination of the theory uncertainties from missing higher orders \\[0.3cm]
  on NNLO parton distributions with percent accuracy }

  \vspace{1.1cm}

  {\bf The NNPDF Collaboration}: \\[0.1cm]
  Richard D. Ball$^1$,
Andrea Barontini$^2$,
Alessandro Candido$^{2,3}$,
Stefano Carrazza$^2$,
Juan Cruz-Martinez$^3$,\\[0.1cm]
Luigi Del Debbio$^1$,
Stefano Forte$^2$,
Tommaso Giani$^{4,5}$,
Felix Hekhorn$^{2,6,7}$,
Zahari Kassabov$^8$,\\[0.1cm]
Niccol\`o Laurenti,$^2$
Giacomo Magni$^{4,5}$,
Emanuele R. Nocera$^9$,
Tanjona R. Rabemananjara$^{4,5}$,
Juan Rojo$^{4,5}$,\\[0.1cm]
Christopher Schwan$^{10}$,
Roy Stegeman$^1$, and
Maria Ubiali$^8$

 \vspace{0.7cm}
 
 {\it \small

 ~$^1$The Higgs Centre for Theoretical Physics, University of Edinburgh,\\
   JCMB, KB, Mayfield Rd, Edinburgh EH9 3JZ, Scotland\\[0.1cm]
 ~$^2$Tif Lab, Dipartimento di Fisica, Universit\`a di Milano and\\
   INFN, Sezione di Milano, Via Celoria 16, I-20133 Milano, Italy\\[0.1cm]
   ~$^3$CERN, Theoretical Physics Department, CH-1211 Geneva 23, Switzerland\\[0.1cm]
    ~$^4$Department of Physics and Astronomy, Vrije Universiteit, NL-1081 HV Amsterdam\\[0.1cm]
   ~$^5$Nikhef Theory Group, Science Park 105, 1098 XG Amsterdam, The Netherlands\\[0.1cm]
~$^6$University of Jyvaskyla, Department of Physics, P.O. Box 35, FI-40014 University of Jyvaskyla, Finland\\[0.1cm]
~$^7$Helsinki Institute of Physics, P.O. Box 64, FI-00014 University of Helsinki, Finland\\[0.1cm]
   ~$^8$ DAMTP, University of Cambridge, Wilberforce Road, Cambridge, CB3 0WA, United Kingdom\\[0.1cm]
     ~$^9$ Dipartimento di Fisica, Universit\`a degli Studi di Torino and\\
   INFN, Sezione di Torino, Via Pietro Giuria 1, I-10125 Torino, Italy\\[0.1cm]
   ~$^{10}$Universit\"at W\"urzburg, Institut f\"ur Theoretische Physik und Astrophysik, 97074 W\"urzburg, Germany\\[0.1cm]

   }

 \vspace{0.7cm}

{\bf \large Abstract}

\end{center}

We include uncertainties due to missing higher order
 corrections to QCD  computations (MHOU) used in the determination of parton
distributions (PDFs) in
the recent NNPDF4.0 set of PDFs. We use our previously
published methodology, based on the treatment of MHOUs and their full
correlations through a theory covariance matrix determined by scale
variation, now fully incorporated in the new NNPDF theory pipeline.
 We assess the impact of the
inclusion of MHOUs on the NNPDF4.0  central values and uncertainties, and
specifically show that they lead to improved consistency of the PDF
determination. PDF uncertainties on physical predictions  in the data
region are consequently either unchanged or moderately reduced by the
inclusion of MHOUs. 
\clearpage
\tableofcontents

\section{Introduction}
\label{sec:intro}

The uncertainty on the parton distribution functions (PDFs) that enter
any prediction of physical processes is one main bottleneck for
precision physics at the LHC. Thanks to methodological progress,
especially the use of machine learning techniques and the 
increase of experimental information, we have recently
achieved a determination of PDFs, NNPDF4.0~\cite{NNPDF:2021njg},
whose nominal precision reaches the percent level. It is clearly crucial to
assess whether this level of precision is reliable, and whether it is matched by the
same level of accuracy.

A considerable effort has gone into assessing the impact on
uncertainties of the methodology used for the determination of PDFs,
and specifically the way it propagates the information contained in
the data onto the PDF  uncertainty (see e.g. the recent studies in
Refs.~\cite{DelDebbio:2021whr,Ball:2022uon}). However, PDF uncertainties, as
given in all standard PDF sets, such as NNPDF4.0~\cite{NNPDF:2021njg},
CT18~\cite{Hou:2019efy}, MSHT20~\cite{Bailey:2020ooq} or
ABMP16~\cite{Alekhin:2017kpj}, do not include theoretical 
uncertainties, i.e., the uncertainties that affect the predictions that are
compared to the data in the process of determining PDFs from a set of experimental data. The only
exceptions are the parametric uncertainty related to the value of the strong
coupling $\alpha_s$, which is routinely included since the early days of LHC
physics~\cite{Demartin:2010er}, and nuclear uncertainties that affect
e.g.\ deep-inelastic scattering (DIS) data on nuclear targets (such as
neutrino DIS data), that are for instance included 
in the NNPDF4.0 PDF determination~\cite{Ball:2018twp,Ball:2020xqw}.

In principle, theoretical uncertainties may come from a variety of different
sources, both parametric (such as the values of the heavy quark masses) and
non-parametric (such as the aforementioned nuclear corrections). Theory
uncertainties related to missing higher orders in QCD computations --- MHOUs
henceforth --- are particularly relevant, because they affect any prediction.
The  current typical perturbative accuracy of QCD computations is
next-to-next-to-leading order
(NNLO), with N$^3$LO corrections only known in a small number of
cases~\cite{Baglio:2022wzu}. At NNLO, MHOUs are  typically
of the order of a few percent or bigger. For LHC precision observables used for
PDF determination, such as gauge boson or top-pair production, this is
surely comparable to the experimental systematic uncertainties, and often
larger or even much larger  than the experimental statistical uncertainty.
Since the uncertainty on the experimental measurement and on
the theoretical prediction enter in a completely symmetric way in the
figure of merit used for PDF determination~\cite{DelDebbio:2021whr},
there is no justification to include the former and not the latter if
they are of comparable sizes.

In Refs.~\cite{NNPDF:2019vjt,NNPDF:2019ubu} we have presented a
methodology for the systematic inclusion of theory uncertainties
in PDF fits through a theory covariance matrix, and for the
computation of the theory covariance matrix related to MHOU through
correlated scale variations. A first application of this methodology to the
construction of a set of NLO PDFs with MHOUs, based on the
NNPDF3.1~\cite{NNPDF:2017mvq} PDF set and methodology, was also
presented in these references, but no global NNLO PDF set with inclusion of
MHOUs is currently available. Such a construction is now greatly
facilitated by the availability of the {\sc\small
  EKO}~\cite{Candido:2022tld} evolution code, and its inclusion in a
new pipeline for producing theory predictions for PDF
determination~\cite{Barontini:2023vmr}, recently used for the
construction of the NNPDF4.0QED PDF set~\cite{NNPDF:2024djq} and currently
adopted by NNPDF as a standard. Other approaches to the determination
of MHOUs have been proposed~\cite{Cacciari:2011ze,David:2013gaa,Bagnaschi:2014wea,Bonvini:2020xeo,Duhr:2021mfd,Kassabov:2022orn}; 
also, MHOUs on NNLO PDFs have been recently estimated based on an
approximate N$^3$LO PDF determination~\cite{McGowan:2022nag}.

It is the purpose of this paper to include MHOUs in the  NNPDF4.0 NLO
and NNLO sets of parton distributions using the methodology of Refs.~\cite{NNPDF:2019vjt,NNPDF:2019ubu}.
The final deliverables of the paper are thus new
versions of the NNPDF4.0 global PDF determination, with more accurate
central values and uncertainties that now also account for the
inclusion of MHOUs in the process of PDF determination. It was indeed
shown in Refs.~\cite{NNPDF:2019vjt,NNPDF:2019ubu} that the main effect
of including MHOUs in PDF determination is to modify central values,
specifically leading to better perturbative convergence.
Parton distributions  with MHOUs included in the PDF uncertainty should 
henceforth become the default choice.

This paper is organized as follows. First, in Section~\ref{sec:theory}
we succinctly review the formalism of Refs.~\cite{NNPDF:2019vjt,NNPDF:2019ubu}
for the determination of a covariance matrix accounting for MHOUs and
its implementation in a PDF fit, specifically referring to its
recent implementation in the {\sc\small EKO} evolution code. In
Section~\ref{sec:benchmarking} we present the MHOU covariance
matrices at NLO and NNLO and validate the NLO covariance matrix
by comparing the estimated MHOUs against the known NNLO results. The main 
deliverables of this paper, namely the NNPDF4.0 NLO and NNLO PDFs with
MHOUs, are presented in Section~\ref{sec:results}, where they are
compared, both in terms of central values and uncertainties,
to their counterparts without MHOUs, thereby assessing the impact of
MHOUs on the consistency of the PDF determination. Finally, in
Section~\ref{sec:summary} we discuss the delivery and usage of the
PDFs with MHOUs and summarize our results.
In an
Appendix we provide explicit
expressions for the missing higher order terms that are generated upon
performing renormalization and
factorization scale variation, and that are used in Section~\ref{sec:theory} to
construct the MHOU covariance matrix.

\section{The MHOU covariance matrix}
\label{sec:theory}

The inclusion of  MHOUs is done by supplementing theory predictions with a 
covariance matrix that accounts for their expected correlated
variation upon inclusion of higher-order corrections. This in turn is
estimated through scale variation. The whole construction is explained in
detail in Refs.~\cite{NNPDF:2019vjt,NNPDF:2019ubu}, to which we refer for a
detailed discussion. In this section we provide a brief summary of the
main aspects of the procedure, with specific reference to its
implementation in the {\sc\small EKO}~\cite{Candido:2022tld} code that,
as mentioned, is part of the new NNPDF theory pipeline~\cite{Barontini:2023vmr}
used in this paper. In particular, for ease of reference, in
this Section we adopt the same notation as in the
{\sc\small EKO} documentation, even though
they depart somewhat from those of Ref.~\cite{Candido:2022tld}.

We first summarize the way perturbative expansions of various
quantities are defined;
we then review the way MHOUs on hard cross sections and
anomalous dimensions can be estimated by mean of scale variation, and we finally
summarize the construction of the
MHOU covariance matrix. The expressions that are needed for
scale variation up
to N$^3$LO are summarized in Appendix~\ref{app:explicit}.

\subsection{Perturbative expansion and factorization}
\label{sec:higherorder}

We start by writing down explicitly the perturbative expansion of an observable
factorized in terms of a hard cross-section and PDFs, with the main
goal of establishing notation. Given this, we only consider the
case of inclusive electroproduction with a single parton species. The
hadronic observable is then a structure function $F(Q^2)$ that
depends on a physical scale $Q^2$, and it is written in terms of a
partonic quantity, the coefficient function
 $C(Q^2)$, perturbatively computed as an expansion in the strong
coupling
\begin{equation}\label{eq:asdef}
      a_s(Q^2)\equiv \frac{\alpha_s(Q^2)}{4\pi}
\end{equation}
and a PDF $f(Q^2)$. In Mellin space we simply have
\begin{equation}
\label{eq:simple}
F(Q^2) = C(Q^2) f(Q^2).
\end{equation}

The partonic coefficient function (for hadronic
processes the partonic cross-section) is expanded perturbatively, and
at N$^k$LO it is given by
\begin{align}
  C\left(a_s(Q^2)\right) = a_s^m(Q^2) \sum_{j=0}^k
  \left(a_s(Q^2)\right)^{j} C_j
  \label{eq:cfexp}
\end{align}
where the LO cross-section is ${\cal O}(a_s^m)$: so for deep-inelastic
structure functions $m=0$ (in the case of $F_2, F_3$) or $m=1$ (in the case of $F_L$), 
for top pair production $m=2$ and so on.

The scale dependence of the strong coupling $a_s(Q^2)$ and of the PDF
$f(Q^2)$ are given by
\begin{align}
    \mu^2\frac{\d a_s(\mu^2)}{\d \mu^2} &= \beta(a_s(\mu^2)) = - \sum_{j=0}^k \left(a_s(\mu^2)\right)^{2+j} \beta_j, \label{eq:betafnc} \\
    \mu^2\frac{\d f(\mu^2)}{\d \mu^2} &= -\gamma\left(a_s(\mu^2)\right).
    f(\mu^2) \label{eq:DGLAP}
\end{align}
At N$^k$LO the beta function and anomalous dimensions are respectively
given by
\begin{align}
    \beta(a_s(\mu^2)) &= - \sum_{j=0}^k \left(a_s(\mu^2)\right)^{2+j} \beta_j, \label{eq:betaexp} \\
   \gamma\left(a_s(\mu^2)\right)&=\sum_{j=0}^k \left(a_s(\mu^2)\right)^{1+j} \gamma_j . \label{eq:DGLAPexp}
\end{align}

The coefficients $\beta_j$ are known up to $k=4$ (N$^4$LO or five
loops)~\cite{Baikov:2016tgj,Chetyrkin:2017bjc,Herzog:2017ohr,Luthe:2017ttg},
while the coefficients $\gamma_j$ are known exactly up to $k=2$ and
approximately for $k=3$ (N$^3$LO or four
loops)~\cite{Davies:2016jie,Moch:2017uml,
Davies:2022ofz,Henn:2019swt,Bonvini:2018xvt,Moch:2021qrk,Soar:2009yh,McGowan:2022nag,Falcioni:2023luc}.
The perturbative expansion of the solutions to \cref{eq:betafnc,eq:DGLAP},
which are needed in order to compute the scale variation terms that we are
interested in, are given explicitly in Appendix~\ref{app:explicit}.

The solution to \cref{eq:DGLAP} can be written in the form of an evolution
kernel operator (EKO) $E(\mu^2 \leftarrow \mu_0^2)$~\cite{Candido:2022tld}
given by
\begin{equation}
    f(\mu^2) = E(\mu^2 \leftarrow \mu_0^2) f(\mu_0^2) = \mathcal P \exp\left(-\int_{\mu_0^2}^{\mu^2} \frac{\d{\mu'}^2}{{\mu'}^2}\, \gamma\left(a_s({\mu'}^2)\right)\right)f(\mu_0^2), \label{eq:EKO}
\end{equation}
where $\mathcal P$ denotes path ordering.
Of course, including the anomalous dimension to N$^k$LO accuracy
yields a PDF $f(Q^2)$ whose scale dependence has a resummed
(next-to-)$^k$-leading-logarithmic (N$^k$LL) accuracy.
The perturbative expansion of this solution is given up to (fixed,
i.e.\ not resummed) order $a_s^3$ in Appendix~\ref{app:explicit}.

\subsection{MHOUs from scale variation}
\label{sec:pert}

Theoretical predictions at hadron colliders depend on two quantities that are
 computed perturbatively: the partonic cross sections or coefficient
functions, Eq.~(\ref{eq:cfexp}), and the anomalous dimensions,
Eq.~(\ref{eq:DGLAPexp}), that determine the scale dependence,
Eq.~(\ref{eq:EKO}), of the PDF. Both quantities can be expressed as a series in the strong
coupling $a_s(Q^2)$, in turn perturbatively given, through the
solution to Eq.~(\ref{eq:betafnc}), in terms of the value of the strong coupling at a
reference scale, typically $a_s(M_Z)$. The MHOU on the predictions is due to the truncation of
these perturbative expansions at a given order.

In principle, if a variable-flavor-number
scheme~\cite{Collins:1998rz,Buza:1996wv} is used, a further MHOU is
introduced by the truncation of the perturbative expansion of the
matching conditions that relate PDFs in schemes with a different number
of active flavors. These uncertainties, especially those related to the
matching at the charm threshold, are very important if one is
interested in PDFs below the charm threshold, such as for instance
when trying to determine the intrinsic charm
PDF~\cite{Ball:2022qks,Ball:2023anf}. However, if one is interested in
precision LHC phenomenology, then physics predictions are produced in a $n_f=5$
scheme, but PDFs are also determined by comparing to data predictions
whose vast majority is computed in the $n_f=5$
scheme. Hence, the matching uncertainties only affect the small amount
of data below the bottom threshold (no data below the charm threshold
are used), and then through the MHOU at the bottom threshold, which is
very small. The MHOU related to the matching conditions are thus
subdominant and we will neglect them here.

We thus focus on MHOUs on the hard cross-sections and anomalous dimensions.
The estimate of these MHOUs from scale variation are obtained by producing
various expressions for a perturbative result to a given accuracy,
that differ by the 
subleading terms that are generated when varying the scale at which the strong
coupling is evaluated. Starting with the coefficient function,
Eq.~(\ref{eq:cfexp}), we thus construct a scale-varied N$^k$LO coefficient
function
\begin{equation}\label{eq:renscalvar}
\bar C(a_s(\mu^2),\rho_r)=
a_s^m(\mu^2) \sum_{j=0}^k
  \left(a_s(\mu^2)\right)^{j} \bar C_j(\rho_r)
\end{equation}
by requiring that
\begin{align}
  \bar C(a_s(\rho_r Q^2),\rho_r)=C(a_s(Q^2))\left[1+{\cal O}(a_s)\right],
    \label{eq:Cbarren}
\end{align}
which fixes the scale-varied coefficients $\bar C_j(\rho_r)$ in terms
of the starting
$C_j$. Explicit expressions are given up to N$^3$LO in Appendix~\ref{app:explicit}. At any given order $C$ and $\bar C$ differ by
subleading terms: their difference is taken as an estimate of
the missing higher orders, and  it may be  used for the construction of a
MHOU covariance matrix, as summarized in Sect.~\ref{sec:thcovmat}.
We refer to this way of estimating   MHOUs on  partonic
cross-sections as renormalization scale variation.

Through the same procedure, we may obtain an estimate of the MHOU on
the anomalous 
dimension, Eq.~(\ref{eq:DGLAPexp}). Namely,  we construct a scale-varied N$^k$LO
anomalous dimension
\begin{equation}\label{eq:factscalvar}
\bar \gamma(a_s(\mu^2),\rho_f)= a_s(\mu^2)
 \sum_{j=0}^k
  \left(a_s(\mu^2)\right)^{j} \bar \gamma_j(\rho_f),
\end{equation}
by requiring that
\begin{align}
  \bar\gamma(a_s(\rho_f Q^2),\rho_f)=\gamma(a_s(Q^2))\left[1+O(a_s)\right],
    \label{eq:gammabarren}
\end{align}
which fixes the coefficients $\bar \gamma_j(\rho_f)$ in terms of
$\gamma_j$: of course, the corresponding expressions are the
same as those obtained by expressing the $\bar C_j(\rho_f)$ in terms of
$C_j$, in the particular case $m=1$. Again the subleading difference
between $\gamma$ and $\bar \gamma$ may be taken as an
estimate of the MHOU on anomalous dimensions. This uncertainty then
translates into a MHOU on the PDF $f(Q^2)$ when this is expressed
through Eq.~(\ref{eq:EKO}) in terms of the PDFs at the
parametrization scale.
We refer to this estimate of the MHOU on the scale dependence of the PDF
as factorization scale variation.

By substituting the scale-varied anomalous dimension $\bar\gamma(\alpha(\mu^2),\rho_f)$ in the expression Eq.~(\ref{eq:EKO}) of
the PDF one can show~\cite{NNPDF:2019ubu} that factorization scale
variation can 
be equivalently performed directly at the level of the PDF, by
defining a scale-varied PDF $\bar f(Q^2,\rho_f)$ whose scale
dependence is given by a scale-varied evolution kernel operator (EKO) $ \bar E(Q^2 \leftarrow \mu^2_0,\rho_f)$:
\begin{align}
\bar f(Q^2,\rho_f) = \bar E(Q^2 \leftarrow \mu^2_0,\rho_f) f(\mu_0^2), \label{eq:schemeb}
\end{align}
and the scale-varied EKO $\bar E$, computed at N$^k$LL, differs by subleading terms from
the original EKO:
\begin{equation}
  \bar E(Q^2 \leftarrow \mu^2_0,\rho_f )=E(Q^2 \leftarrow
  \mu^2_0)\left[1+O(a_s)\right]\label{eq:ekoscale}.
\end{equation}
The scale-varied EKO can be constructed as
\begin{equation}
 \bar E(Q^2 \leftarrow \mu^2_0,\rho_f )=K\left(a_s(\rho_f Q^2),\rho_f\right) E(\rho_f Q^2 \leftarrow \mu^2_0),\label{eq:ekok}
\end{equation}
where at N$^k$LL (i.e.\ with the anomalous dimension computed at
N$^k$LO) the additional evolution kernel $K(a_s(\rho_f Q^2),\rho_f)$ is
given by the expansion
\begin{equation}
K\left(a_s(\rho_f Q^2),\rho_f\right)= \sum_{j=0}^{k} \left(a_s(\rho_f Q^2)\right)^{j}K_j(\rho_f)\label{eq:ks}.
\end{equation}
Substituting this expansion in Eq.~(\ref{eq:ekoscale}) fixes all
coefficients $K_j(\rho_f)$ in terms of
$\gamma_j$. Their expressions are given up to N$^3$LO in Appendix~\ref{app:explicit}.
Taken together, \cref{eq:ekoscale,eq:ekok}, mean that the scale-varied evolution kernel Eq.~(\ref{eq:ekoscale})
evolves from $\mu_0^2$ to $\rho_f Q^2$, and then from $\rho_f Q^2$ back
to $Q^2$, but with the latter evolution expanded out to fixed
N$^k$LO.

The two  ways of performing factorization scale variation, on
anomalous dimensions Eqs.~(\ref{eq:factscalvar}-\ref{eq:gammabarren})  or on
PDFs  Eqs.~(\ref{eq:schemeb}-\ref{eq:ekoscale}) are equivalent, as when performed at N$^k$LO they generate the same
subleading N$^{k+1}$LO terms  (though yet higher order terms are
different), namely
\begin{equation}
  \bar f(Q^2,\rho_f)-f(Q^2)= \left\{\mathcal P
  \exp\left(-\int_{\mu_0^2}^{\mu^2} \frac{\d{\mu'}^2}{{\mu'}^2}\,
  \left[\bar\gamma\left(a_s({\rho_f\mu'}^2),\rho_f\right)-\gamma\left(a_s({\mu'}^2)\right)\right]\right)f({\mu'}^2)\right\}\left[1+{\cal O}(a_s)\right].
  \label{eq:avsb}
\end{equation}

These two different ways of performing factorization scale variation, by
varying the scale of the anomalous dimension or varying the scale of
the PDF were respectively referred to as scheme A and scheme B in
Ref.~\cite{NNPDF:2019ubu}. A third way, referred to as scheme C in
Ref.~\cite{NNPDF:2019ubu}, consists of using the scale-varied PDF,
Eqs.~(\ref{eq:schemeb}-\ref{eq:ekok}), namely
\begin{align}
\bar f(Q^2,\rho_f) = K(a_s(\rho_f Q^2),\rho_f) E(\rho_f Q^2 \leftarrow
 \mu^2_0 ) f(\mu_0^2) \label{eq:schemebtoc}
\end{align}
in the factorized expression, Eq.~(\ref{eq:simple}), but including
$K(a_s(\rho_f Q^2),\rho_f)$ in the coefficient function instead of the PDF. This amounts
to evaluating the PDF at a different scale, but with a modified
coefficient function. The corresponding
explicit expressions are also given for completeness in Appendix~\ref{app:explicit}.

In standard practice, factorization scale variation is usually performed using
scheme C, because this does not require changing the PDFs, which are typically
taken as given from an  external provider. However, in the context of a PDF
determination, factorization scale variation through scheme B,
Eq.~(\ref{eq:schemeb}), is simplest, as it only requires
modifying the EKO used to compute PDF evolution. This is the way we
will perform factorization scale variation in this paper.

\subsection{Construction of the covariance matrix}
\label{sec:thcovmat}

The MHOU due to the perturbative truncation of the partonic
cross-sections and the scale dependence of the PDFs, respectively
estimated through renormalization scale variation, \cref{eq:renscalvar,eq:Cbarren}, and
factorization scale variation according to scheme B of Ref.~\cite{NNPDF:2019ubu},
\cref{eq:schemeb,eq:ks}, are included through a MHOU covariance
matrix. This is constructed as
follows~\cite{NNPDF:2019vjt,NNPDF:2019ubu}.

First, we define the shift in theory prediction for the $i$-th datapoint due
to renormalization and factorization scale variation
\begin{equation}\label{eq:delta}
    \Delta_{i}(\rho_{f}, \rho_{r}) \equiv T_{i}(\rho_f, \rho_r) - T_{i}(0, 0),
\end{equation}
where $T_{i}(\rho_f, \rho_r)$ is the prediction for the $i$-th
datapoint obtained by varying the renormalization and factorization
scale by a factor $\rho_r$, $\rho_f$ respectively. Note that in
Refs.~\cite{NNPDF:2019vjt,NNPDF:2019ubu} the scale variations were
parametrized by their logarithms, i.e.\ through parameters
$\kappa_r=\ln\rho_r$, $\kappa_f=\ln\rho_f$.

Next, we choose a correlation pattern for scale variation, as
follows~\cite{NNPDF:2019vjt,NNPDF:2019ubu}:
\begin{itemize}
  \item factorization scale variation is correlated for all datapoints, because the scale dependence of PDFs is universal;
  \item renormalization scale variation is correlated for all
    datapoints belonging to the same category, i.e.\ either the same
    observable (such as, for
    instance, fully inclusive DIS cross-sections) or to
    different observables for the same process (such as, for
    example, the $Z$ transverse momentum and rapidity distributions).
\end{itemize}
Note that this requires a categorization of processes: for instance we
consider charged-current and neutral-current deep-inelastic scattering
as separate processes. The particular process categorization adopted in this work
is discussed in Section~\ref{subsec:dataset}.

These choices correspond to the assumptions that factorization and
renormalization scale variation fully capture the MHOU on anomalous
dimensions and partonic cross-sections respectively, and that missing
higher order terms are of a similar nature and thus of a similar size
in all processes included in a given process category.
Different assumptions are consequently possible, for
instance decorrelating the renormalization scale variation from
contributions to the same process from different partonic
sub-channels, or introducing a further variation of the scale of the
process on top of the renormalization and factorization scale
variation discussed above (see Section~4.3 of Ref.~\cite{NNPDF:2019ubu}  for a more
detailed discussion).

We then define a MHOU covariance matrix, whose matrix element  between two datapoints
$i$, $j$ is
\begin{equation}
\label{eq:thcovmat}
    S_{ij} = n_{m}\sum_{V_{m}} \Delta_{i}(\rho_{f}, \rho_{r_{i}})\Delta_{j}(\rho_{f}, \rho_{r_{j}}),
\end{equation}
where the sum runs over the space $V_m$ of the $m$ scale variations
that are included; the factorization scale $\rho_f$ is always varied
in a correlated way, the renormalization scales $\rho_{r_i}$, $\rho_{r_j}$ are varied in a
correlated way ($\rho_{r_i}=\rho_{r_j}$) if datapoints $i$ and $j$ belong to
the same category, but are varied independently if $i$ and $j$
belong to different categories, and $n_m$ is a normalization
factor. The computation of the normalization factor is nontrivial
because it must account for the mismatch between the dimension of the
space of scale variations when two datapoints are in the same category
(so there is only one correlated set of renormalization scale
variations) and when they are not (so there are two independent sets
of variations). These normalization factors were computed for various
choices of the space $V_m$ of scale variations and for various values
of $m$ in Ref.~\cite{NNPDF:2019ubu}, to which we refer for details.

As in Refs.~\cite{NNPDF:2019vjt,NNPDF:2019ubu} we consider scale variation by a factor 2, so
\begin{equation}\label{eq:scales}
  \kappa_f=\ln \rho_f=\pm\ln 4\qquad
  \kappa_r=\ln \rho_r=\pm\ln 4.
\end{equation}
In Ref.~\cite{NNPDF:2019ubu} various different choices for the space
of allowed variations were considered. These included, among others: the 9-point
prescription, in which $\kappa_r$, $\kappa_f$ are allowed to both take all
values $(0,\pm\ln 4)$, with
$m=8$ (eight variations about the central value); and 
the commonly used 7-point prescription, with $m=6$, which is obtained from the
former by discarding the two outermost variations, in which
$\kappa_r=+\ln 4$, $\kappa_f=-\ln 4$ or $\kappa_r=-\ln 4$,
$\kappa_f=+\ln 4$. We will show in
Section~\ref{sec:benchmarking} that, upon validation of the MHOU
covariance matrix, the 7--point and 9-point prescription have a
similar behavior, in agreement with what was already found in
Ref.~\cite{NNPDF:2019ubu}. Other prescriptions, with a more limited set of
independent scale variations, where shown in Ref.~\cite{NNPDF:2019ubu}
to perform less well, and we will not consider them any further.
The explicit expressions for the MHOU covariance matrix with the
7-point and 9-point prescription are respectively given in
Eqs.~(4.18-4.19)  and Eq.~(4.15) of Ref.~\cite{NNPDF:2019ubu}. 

The set of assumptions that include the correlation patterns of
renormalization and factorization scale variations, the process
categorization, the range of variation of the scales, and the specific
choice of  variation points involves a certain degree of arbitrariness.
This is inevitable given that the MHOU is the  estimate of
the probability distribution for the
size of an unknown quantity which has an unique true value, and
thus it is intrinsically Bayesian. The only way to validate this kind of estimate is by
comparing its performance to cases in which the true value is known,
as we shall do in Section~\ref{sec:validation}.

\section{The MHOU covariance matrix and its validation}
\label{sec:benchmarking}

We now compute and validate MHOU covariance matrix obtained using the procedure
discussed in the previous section: first, we present its construction
based on a suitable dataset categorization, and then its validation at
NLO where the next-order corrections are known, so the true MHOUs can
be determined exactly.

\subsection{The covariance matrix at NLO and NNLO}
\label{subsec:dataset}

In order to determine the covariance matrix we must
first choose a dataset and process categorization. 
The dataset used for the determination of the NNPDF4.0MHOU PDFs is the same as
the dataset used for the determination of the NNLO NNPDF4.0 PDFs, see
Ref.~\cite{NNPDF:2021njg} for details.
This same dataset is adopted both for the NLO and NNLO  NNPDF4.0MHOU PDFs
discussed here. In Ref.~\cite{NNPDF:2021njg} a somewhat different
dataset was used for the NLO PDF determination, in particular
excluding datapoints for which NNLO corrections are sizable and
including at NLO some data for which NNLO corrections were not
available at the time of the writing of that paper.

Here we wish to adopt
exactly the same dataset at NLO and
NNLO in order to be able to analyze the impact of the inclusion of
MHOUs on perturbative convergence without changes in dataset acting as
a confounding effect. Note that this involves first, including in the
NLO dataset also datapoints for which NNLO corrections are known to be
very large, and furthermore including in both datasets datapoints for which
downward scale variation leads to a low scale. The use of a
MHOU covariance matrix should  take care of both issues. Indeed,
data with  large NNLO corrections should have
correspondingly large MHOUs, to the extent that the estimate based
on scale variation is accurate. Also, scale variation of low scale
data in the worst case  will induce  sizable shifts and thus large 
MHOUs that will possibly reduce the constraining power of these data
in the direction of the shift, thereby effectively deweighting the
data.

As explained in the previous Section, process categories correspond to
classes of processes for which missing higher order terms are likely
to be of similar enough origin. Therefore the correlation between the MHOU
on any pair of predictions for processes in the same category can be
computed to good approximation as if they were two datapoints for the
same physics process. We thus group processes into nine process
categories, namely, neutral-current deep-inelastic scattering (DIS NC);
charged-current deep-inelastic scattering (DIS CC); and the following
seven hadronic production processes: top-pair; $Z$, i.e.\ neutral-current
Drell-Yan (DY NC);  $W^\pm$, i.e.\ charged current Drell-Yan  (DY CC); single top;
single-inclusive jets; prompt photon; dijet.

\begin{figure}[!t]
  \centering
  \begin{subfigure}[!t]{0.49\textwidth}
    \centering
    \includegraphics[width=\textwidth]{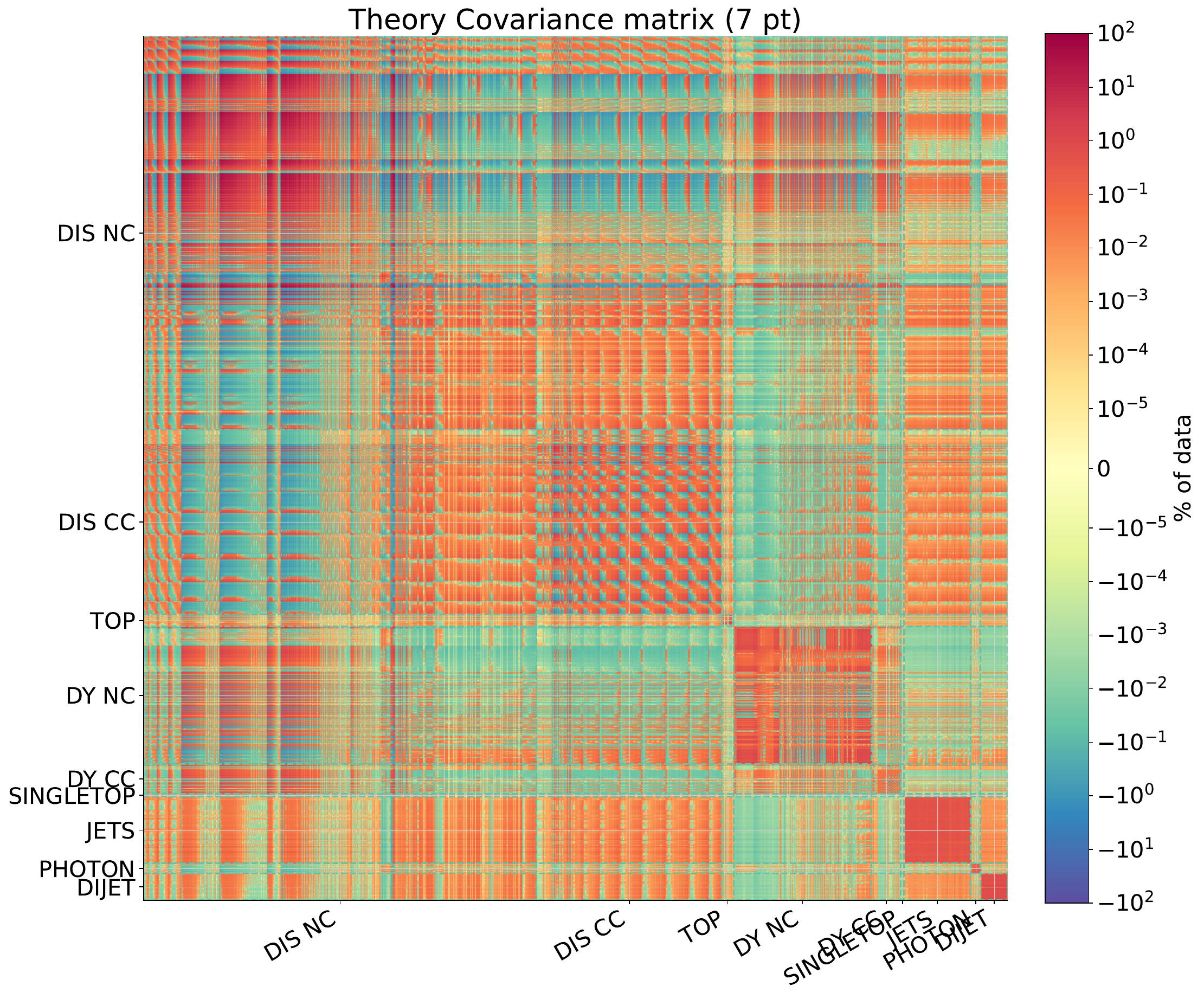}
    \caption{NLO}
  \end{subfigure}
  \hfill
  \begin{subfigure}[!t]{0.49\textwidth}
    \centering
    \includegraphics[width=\textwidth]{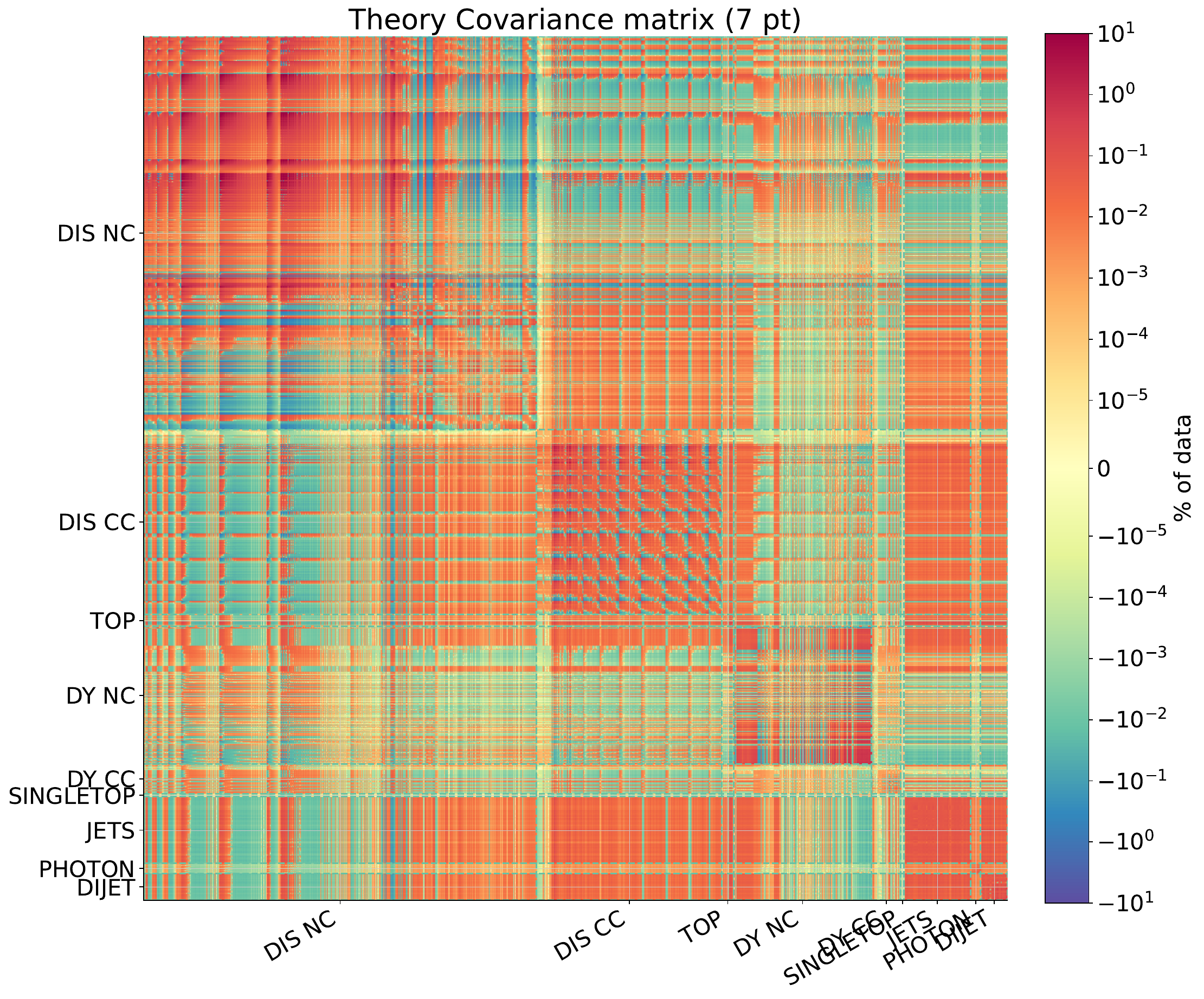}
    \caption{NNLO}
  \end{subfigure}
  \caption{The MHOU covariance matrix Eq.~(\ref{eq:thcovmat})
    computed with the 7-point prescription at NLO (left) and NNLO
    (right). Note that range of the color scale is by one order of
    magnitude wider in the NLO case.}
  \label{fig:thcovmats}
\end{figure}

With these choices the covariance matrices at NLO and NNLO can be computed from
Eq.~(\ref{eq:thcovmat}). Results at NLO and NNLO computed using the
7-point prescription are shown in Fig.~\ref{fig:thcovmats}. As expected,
the absolute value of the matrix elements is almost always
smaller at NNLO than at NLO:
the reduction is in fact typically by more than a factor 2 and on
average almost one order of magnitude. However, the pattern
of correlations appears to be quite stable upon changes of perturbative
order. Note that all data points, including those that belong to different
experiments, are correlated through MHOUs on perturbative evolution. This is
a significant difference in comparison to a typical experimental covariance
matrix.

\begin{figure}[!t]
  \centering
  \begin{subfigure}[!t]{\textwidth}
    \centering
    \includegraphics[width=\textwidth]{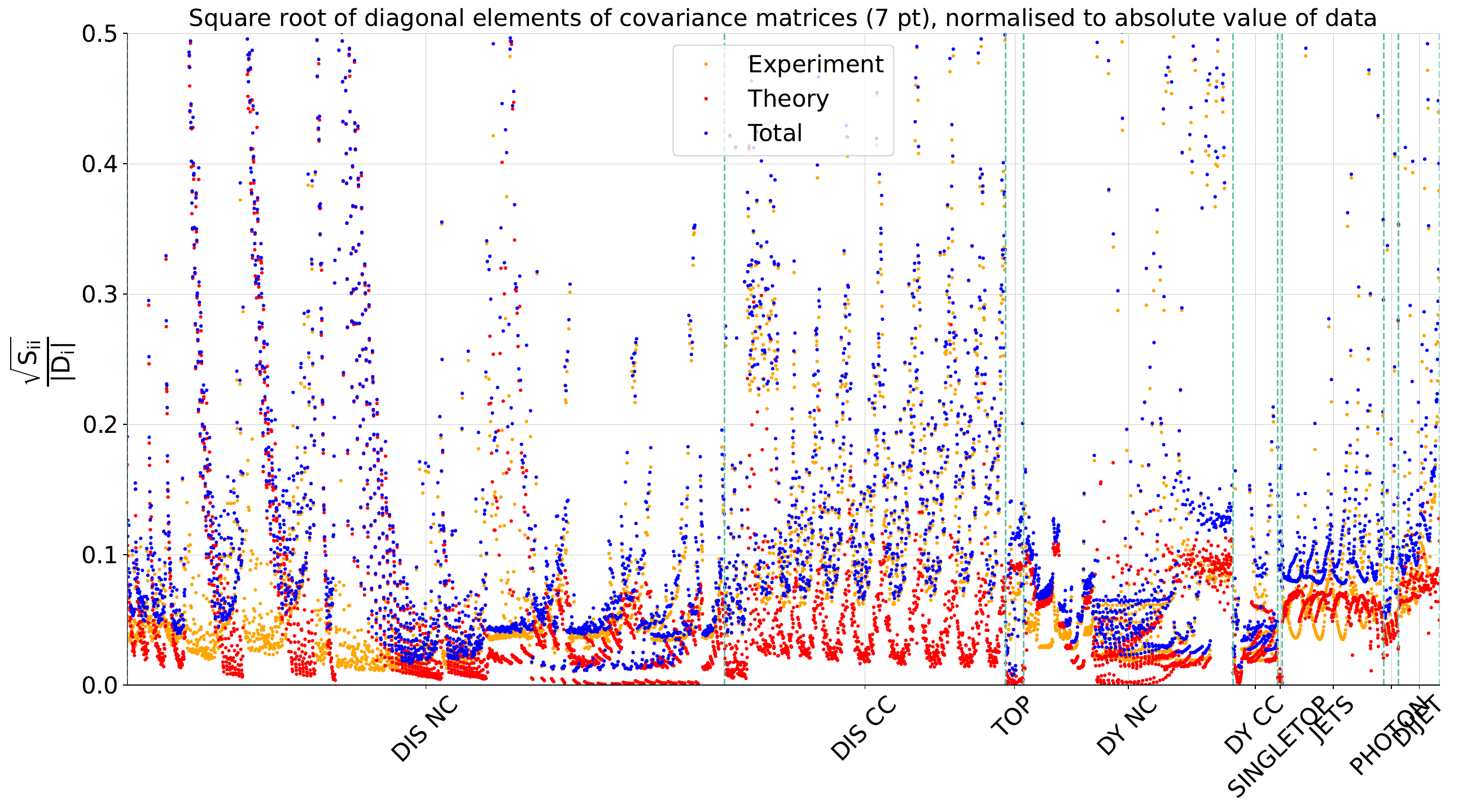}
    \caption{NLO}
  \end{subfigure}
  \begin{subfigure}[!t]{\textwidth}
    \centering
    \includegraphics[width=\textwidth]{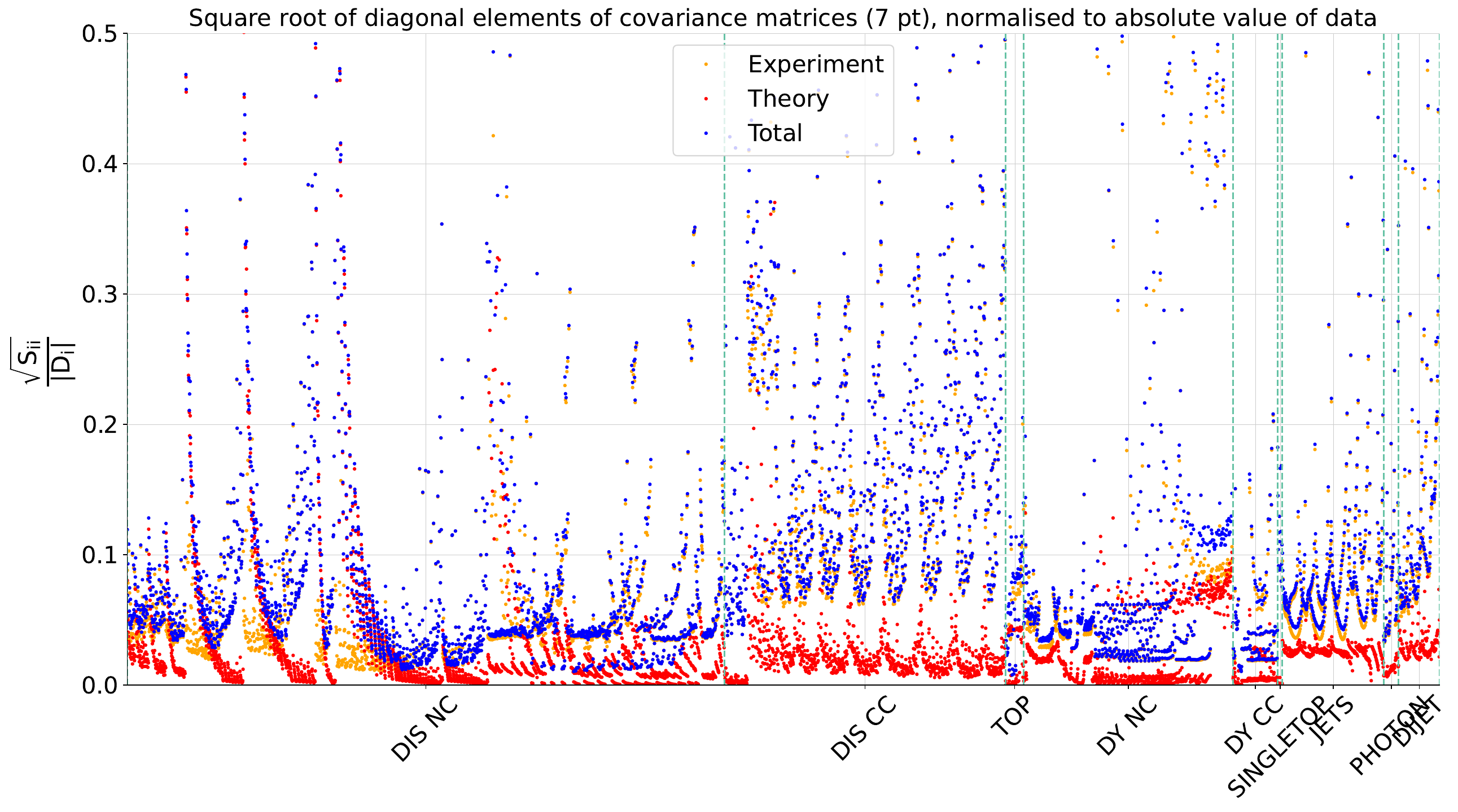}
    \caption{NNLO}
  \end{subfigure}
  \caption{Comparison of the experimental and MHO contributions to the
    relative uncertainty, defined as the square root of the diagonal
    element of the covariance matrix
    normalized to the value of the theory prediction, at NLO (top) and NNLO
    (bottom), for all datapoints. The experimental,  MHO, and total
    uncertainties are shown in yellow, red and blue respectively.}
  \label{fig:diag_elmn}
\end{figure}

The relative uncertainties on individual points (i.e.\ the square-root of the
diagonal covariance matrix elements) before and after the inclusion
of the MHOU and the MHOU itself are
compared in Fig.~\ref{fig:diag_elmn} at NLO and NNLO. 
It is clear that for hadronic processes
at NLO the MHOU uncertainty is on average the same size as the
experimental uncertainty, while at NNLO
it is subdominant. For DIS the difference between NLO and NNLO
is less marked, but at NNLO the uncertainty is again subdominant, except
at small $x$ and $Q^2$.  Consequently, we might expect the effect of
MHOUs at NNLO to be mostly through correlations, and thus to mostly
impact PDF central values, and less so PDF uncertainties, except at
small $x$, for the PDF combinations that dominate the small-$x$
behavior, i.e. gluon and singlet.

\subsection{Validation}
\label{sec:validation}

The MHOU covariance matrix at NLO can be validated by comparing it to
the known difference between NLO and NNLO predictions. This
comparison can be performed using various estimators,  originally
proposed in Ref.\cite{NNPDF:2019vjt}.  
We present here results of this validation, both for our default
7-point prescription, as well as for the 9-point prescription
discussed in Section~\ref{sec:thcovmat}.

We define a normalized shift vector, whose $i$-th component   $\delta_{i}$
is the normalized shift of the $i$-th datapoint due to the change in theory
prediction from NLO to NNLO for fixed PDF, namely
\begin{equation}\label{eq:shift}
  \delta_{i} = \frac{T_{i}^{\text{NNLO}} - T_{i}^{\text{NLO}}}{T_{i}^{\text{NLO}}},
\end{equation}
where the index $i$ runs over all the datapoints, and $T_{i}^{\text{NNLO}}$ and 
$T_{i}^{\text{NLO}}$ are respectively the NNLO and NLO theory
predictions both computed using the NLO PDF set. 

\begin{figure}[!t]
  \centering
  \begin{subfigure}[!t]{\textwidth}
    \centering
    \includegraphics[width=\textwidth]{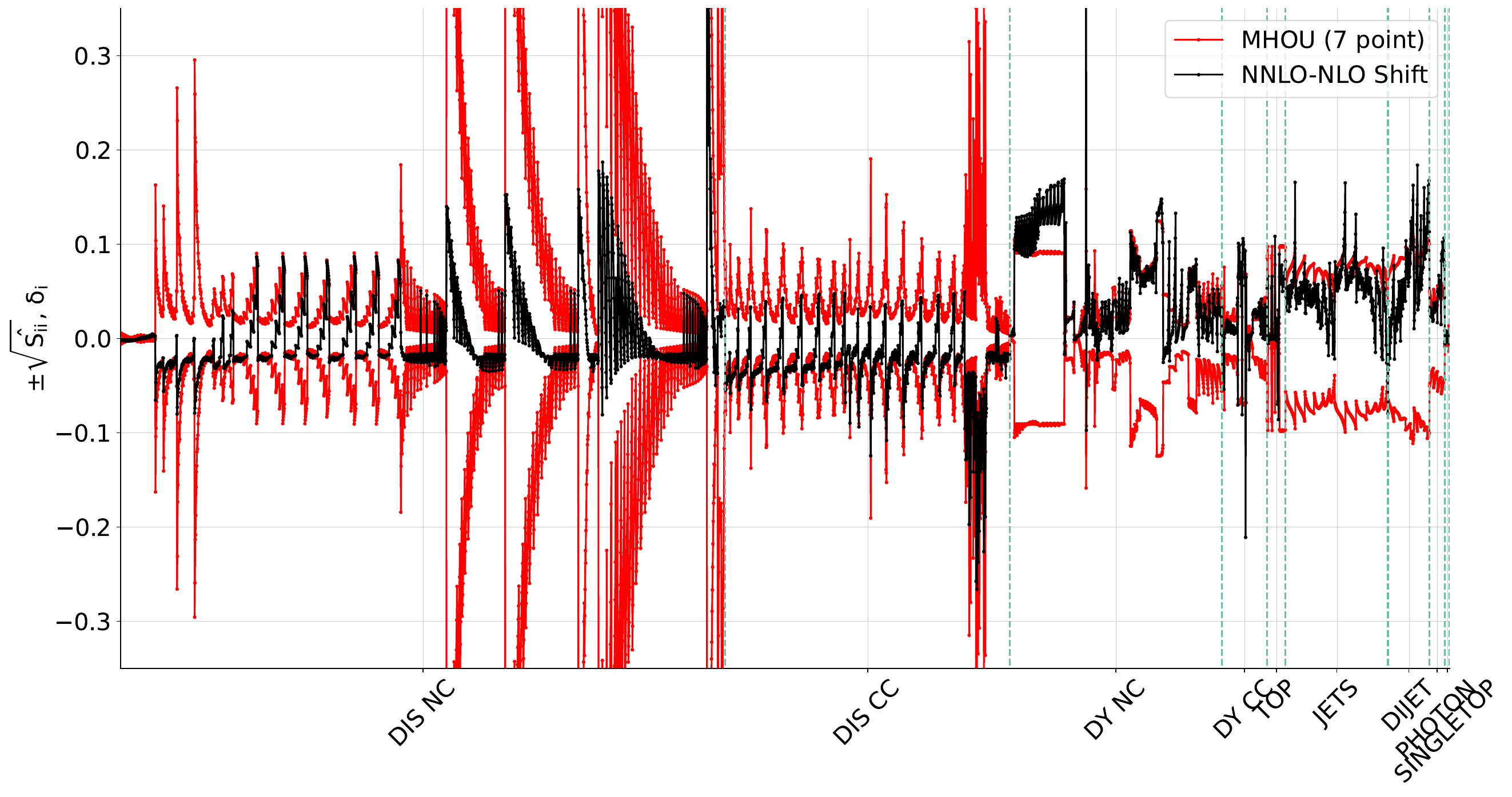}
    \caption{7 point}
  \end{subfigure}
  \begin{subfigure}[!t]{\textwidth}
    \centering
    \includegraphics[width=\textwidth]{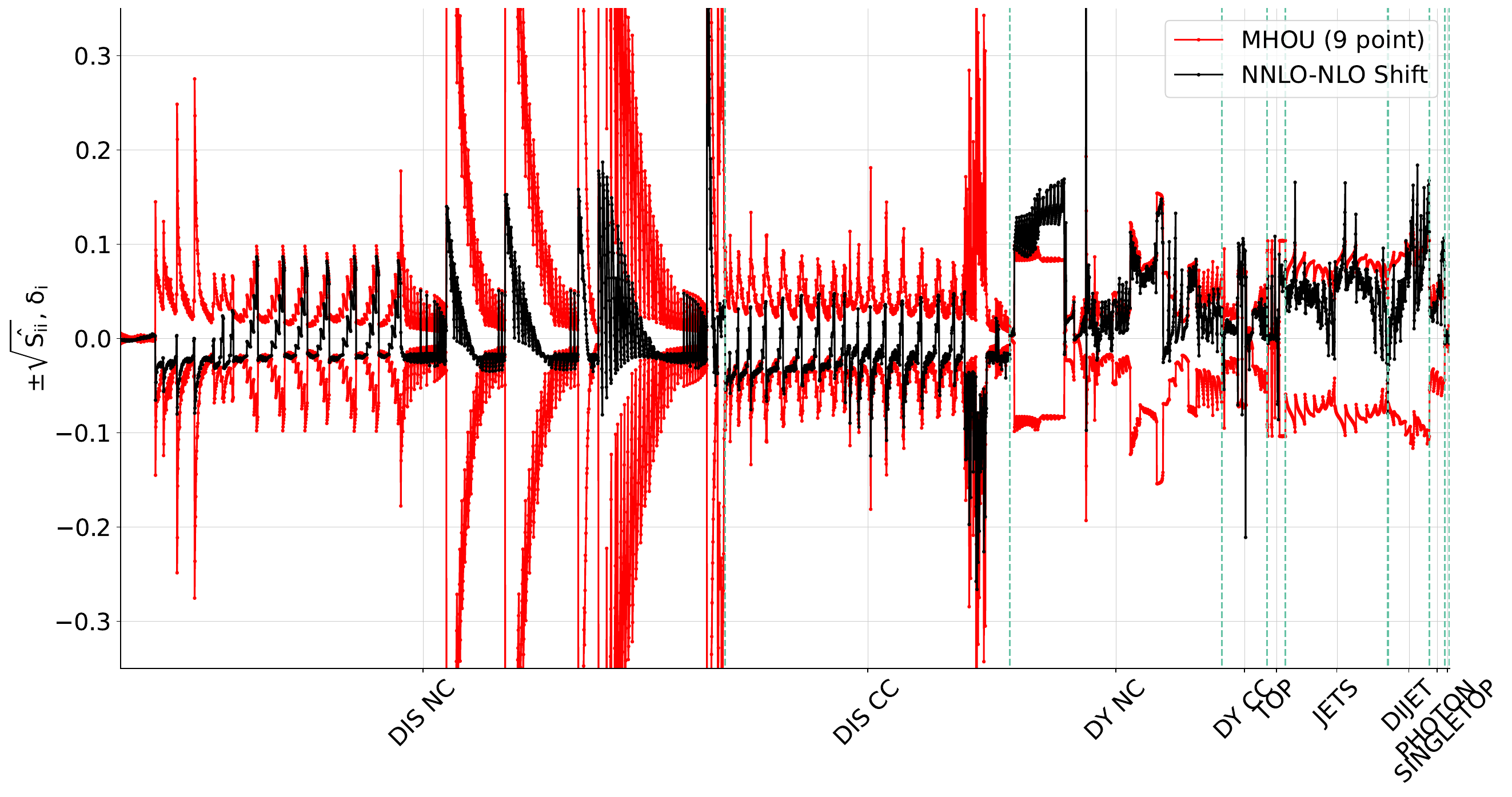}
    \caption{9 point}
  \end{subfigure}
  \caption{Comparison of the symmetrized NLO MHOU  $\pm{\sqrt{\hat{S}_{ii}}}$
    (red, same as in
    Fig.~\ref{fig:diag_elmn}, but normalized to the NLO theory
    prediction)  to the normalized NNLO-NLO shift $\delta_i$
    Eq.~(\ref{eq:shift}) (black) for all 
    datapoints. Results obtained with the 7-point (top) and 9-point
    (bottom) prescription are shown.}
  \label{fig:shift_diag}
\end{figure}

The simplest validation consists of comparing  the shift $\delta_{i}$ to  
the uncertainty on individual points (also normalized), i.e.\ to the square
root of the diagonal entries of the normalized NLO  MHOU covariance matrix 
\begin{equation}\label{eq:normcov}
  \hat{S}_{ij}^{\text{NLO}} = \frac{S_{ij}^{\text{NLO}}}{T_{i}^{\text{NLO}}T_{j}^{\text{NLO}}}\, . 
\end{equation}
Results are shown, for both the 7-point and the
9-point prescriptions, in Fig.~\ref{fig:shift_diag}, where we compare $\delta_i$ to $\pm\sqrt{\hat{S}_{ii}}$. It is clear that for DIS both 7-point and 9-point
scale variations at NLO provide a very conservative
uncertainty estimate that significantly overestimates the NLO-NNLO
shift. On the other hand for hadronic processes the shift and scale variation
estimate are generally comparable in size. Only for DY scale
variations perform less well, with instances of underestimation of
the shift. Whereas this may suggest adjusting the range of scale
variation on a process-by-process basis, it is unclear to which extent the NLO
behavior could be generalized to higher orders: perhaps this approach
could be pursued once more orders are known, along the lines of Refs.~\cite{Cacciari:2011ze,Bagnaschi:2014wea,Bonvini:2020xeo,Duhr:2021mfd}.

This validation  is however very crude, in that it does not test correlations
at all. These can be checked by comparing the eigenvalues of the
covariance matrix to the projection of the shift along its eigenvectors.
It is important to realize that the shift  $\delta_{i}$ is a vector
in the $N_{\rm dat}$-dimensional space of data, of which the 
independent eigenvectors  of the covariance matrix span a small
subspace $S$ with dimension $N_{\rm sub}\ll N_{\rm dat}$. In our case,
$N_{\rm dat}=4616$ while $N_{\rm sub}=22$
for 7-point scale variation, and  $N_{\rm sub}=48$
for 9-point (see formulae in Appendix~A of Ref.~\cite{NNPDF:2019vjt} with
$p=9$ process classes). Therefore, a further nontrivial requirement
is that the shift vector be mostly contained within the subspace $S$.

We can perform both tests quantitatively as follows~\cite{NNPDF:2019vjt}.
First, we determine the eigenvectors $e_{i}^{\alpha}$ and the
eigenvalues $\lambda^{\alpha} = (s^{\alpha})^{2}$ of the MHOU
covariance matrix, with $s^{\alpha}>0$. Then, we determine 
the $N_{\rm sub}$  projections $\delta^\alpha$ of the shift vector $\delta_{i}$
on the eigenvectors $e_{i}^{\alpha}$, i.e.\
\begin{equation}
  \label{eq:projection}
  \delta^{\alpha} =
  \sum_{i=1}^{N_{\text{dat}}}\delta_{i}e_{i}^{\alpha}, \quad
  \alpha=1,\dots,N_{\rm  sub}.
\end{equation}
Finally, we determine
the component of the shift vector in the $N_{\rm sub}$ dimensional
subspace $S$:
\begin{equation}
  \delta^{S}_i = \sum_{\alpha=1}^{N_{\rm sub}} \delta^{\alpha}
  e_{i}^{\alpha}, 
\end{equation}
and the orthogonal component
\begin{equation}\label{eq:miss}
  \delta^{\rm miss}_i=\delta_i-\delta^{S}_i,
\end{equation}
which is the part of the shift vector that is missed by the MHOU
covariance matrix.

We can now test whether correlated uncertainties are correctly
accounted for by checking whether $s^\alpha$ are of comparable size of
$\delta^\alpha$ --- in principle, assuming MHO terms to be Gaussianly
distributed, 68\% of $\delta^\alpha$ should be smaller than or equal
to  $s^\alpha$. We can further test how  much of the shift vectors
lies in the subspace $S$ by determining the length $|\delta^{\rm miss}|$ of the
missed vector, and the angle between the full shift
vector and its component contained in the $S$ subspace
\begin{equation}
\label{eq:theta}
  \theta =
  \arccos{\left(\frac{|\delta^{S}|}{|\delta|}\right)}.
  \end{equation}
Clearly, if the shift was entirely explained by the MHOU covariance
matrix, then $|\delta^{\rm miss}|=0$, $|\delta^{S}|=|\delta|$, and $\theta=0$.

Related to this, it is interesting to observe that the way
scale variation prescriptions affect the final result significantly
differs when using a MHOU  covariance matrix approach, in
comparison to the frequently adopted method of estimating MHOUs by
taking the envelope of results that are found when varying the
scales (as e.g. discussed in Sect.~12.4 of
Ref.~\cite{LHCHiggsCrossSectionWorkingGroup:2011wcg}). In the latter case,
if the shift produced by scale variation is for some reason
unnatural, taking the envelope  may lead to overestimated uncertainties. This is in fact
the standard argument for favoring the 7-point prescription over the 9-point
prescription, as the latter may generate unnaturally large scale
ratios. In contrast, with a  covariance matrix formalism, an unnatural
shift corresponds to a large eigenvalue associate to an eigenvector
along which the actual shift is instead small, or perhaps even
zero. The effect of this is generally moderate or innocuous: the large
eigenvalue  means that the best fit can move in the direction of the
corresponding eigenvector at little cost in $\chi^2$, but if the actual shift
is small, nothing should be gained by moving in that
direction. 

\begin{figure}[!t]
  \centering
  \begin{subfigure}[!t]{0.49\textwidth}
    \centering
    \includegraphics[width=\textwidth]{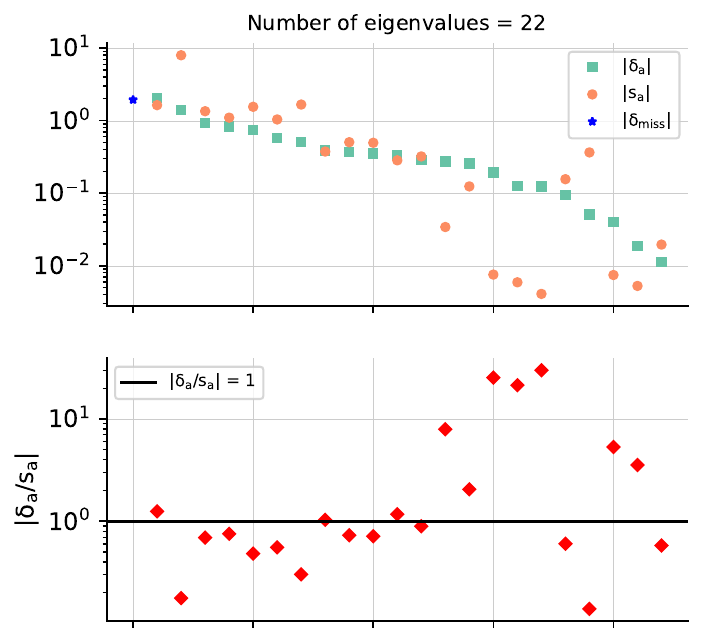}
    \caption{7-point}
  \end{subfigure}
  ~
  \begin{subfigure}[!t]{0.49\textwidth}
    \centering
    \includegraphics[width=\textwidth]{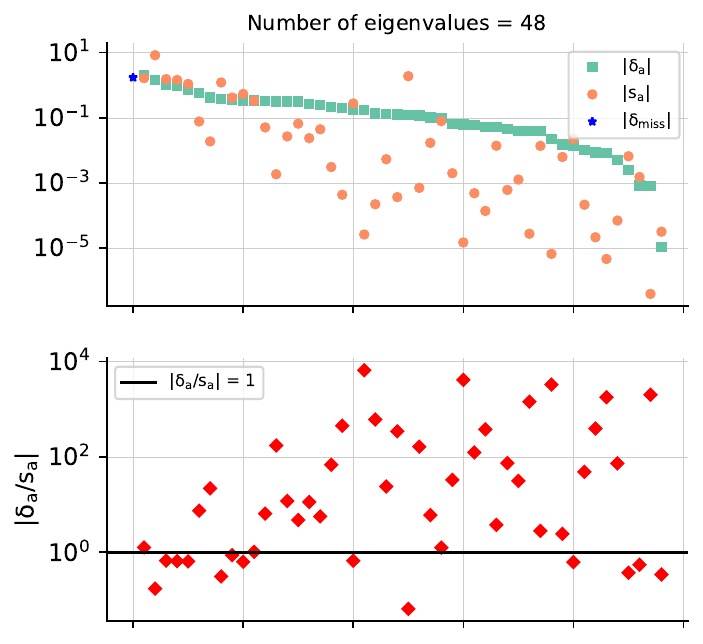}
    \caption{9-point}
  \end{subfigure}
  \caption{The projection $|\delta^{\alpha}|$
    Eq.~\ref{eq:projection} of the
    shift vector along each eigenvector of the MHOU covariance matrix
    compared to the square root $|s^{\alpha}|$ of the corresponding eigenvalue,
    for both the 7- and the 9-point prescription. The plots are ordered by
    decreasing size of the projections and the results are shown both as
    absolute and as a ratio. In the absolute panel, the length
    $|\delta^{\text{miss}}|$ Eq.~(\ref{eq:miss}) of
    the missed component Eq.~(\ref{eq:miss}) is also shown.}
  \label{fig:projection}
\end{figure}

Consequently, what matters in judging the effectiveness of the theory 
covariance matrix is whether the largest components of the shift
vector are well reproduced by the corresponding
theory covariance matrix eigenvalues (and specifically not
underestimated). In practice, we order the 
shift projections $|\delta^{\alpha}|$ by decreasing size, and we
compare them in
Fig.~\ref{fig:projection}  to the covariance matrix
eigenvalues $|s^{\alpha}|$,  
both for the 7-point and the 9-point prescriptions. We also show in figure  the
length of the 
missed component $|\delta^{\rm miss}|$. There is good
agreement between shift projections and predicted MHOUs for the
largest eigenvectors  using both prescriptions.
For smaller eigenvectors there is also generally good 
agreement, but the 9-point prescription somewhat underestimates the
size of individual components of the shift. 

\begin{table}[!t]
  \scriptsize
  \centering
  \renewcommand{\arraystretch}{1.4}
  \begin{tabularx}{\textwidth}{Xcrrrrrrrrrr}
  &
  & \rotatebox{90}{DIS NC}
  & \rotatebox{90}{DIS CC}
  & \rotatebox{90}{TOP}
  & \rotatebox{90}{DY NC}
  & \rotatebox{90}{DY CC}
  & \rotatebox{90}{SINGLETOP}
  & \rotatebox{90}{JETS}
  & \rotatebox{90}{PHOTON}
  & \rotatebox{90}{DIJET}
  & \rotatebox{90}{TOTAL}\\
  Prescription & $N_{\rm sub}$ & \multicolumn{10}{c}{$\theta$ [$\degree$]} \\
  \toprule
  7-point & 22 & 39 & 18 & 24 & 23 & 38 & 14 & 15 & 12 & 12 & 32 \\
  9-point & 48 & 37 & 15 & 20 & 23 & 34 & 12 & 13 &  7 & 12 & 28 \\
  \bottomrule
\end{tabularx}

  \caption{The angle $\theta$, Eq.~(\ref{eq:theta}), for the 7-point and 
    the 9-point prescriptions for individual process categories and for the
    total dataset. The dimension $N_{\rm sub}$ of the $S$ subspace is also shown.}
  \label{tab:angles}
\end{table}

The size of the missed component of shift vector can be seen in 
Fig.~\ref{fig:projection} to be similar to its largest eigenvector
component for both prescriptions, i.e.\ relatively small in comparison
to the full shift, given that the first ten components or so are of comparable
size. Indeed,  this can be seen from the 
angle Eq.~(\ref{eq:theta}) between the shift vector and its
projection in the subspace 
$S$, which  is tabulated in Tab.~\ref{tab:angles} both for individual datasets
and the full dataset. The two prescriptions perform both surprisingly
well, given the very small size of the $S$ subspace, with a very small
difference between the two prescriptions despite the difference by more
than a factor 2 in the size of the $S$ subspace. For almost all  datasets
the direction of the shift and its projection in the $S$ subspace are
quite close and in some cases very close, the only exception being NC
DIS and to a lesser extent DY, especially CC.

In summary, we conclude that the NLO MHOU covariance matrix accounts
quite well for the uncertainty due to the missing NNLO corrections,
with only the uncertainty on the  DY prediction somewhat
underestimated by scale variation, and  the performance of the 7-point
and 9-point prescription fairly 
close. Specifically, the 9-point
prescription performs slightly better in terms of capturing the
subspace in which the shift lies, while the 7-point performs somewhat
better in terms of correctly describing the size of the uncertainty in
the subspace. In the sequel we will adopt the 7-point prescription as
a default.

\section{The NNPDF4.0MHOU determination}
\label{sec:results}

We now turn to the main deliverables of this paper, namely the
NNPDF4.0 NLO and NNLO PDF sets with MHOUs, which are obtained by
repeating the corresponding NNPDF4.0 PDF determinations,
but now also including a MHOU
covariance matrix determined with a
7-point prescription, as discussed in Section~\ref{sec:theory}. The
underlying dataset is identical to that used for the determination of
the NNPDF4.0 NNLO PDFs~\cite{NNPDF:2021njg}. As already mentioned in 
Section~\ref{subsec:dataset}, here we adopt exactly the same dataset at
NLO and NNLO, while in Ref.~\cite{NNPDF:2021njg} a somewhat different
dataset was used at NLO. Hence, we compare here
four PDF sets: NLO and NNLO, with and without MHOUs, all determined based on the
same underlying data.

Note that the NLO PDFs without MHOUs shown here are unsuitable for
phenomenology, 
because they include data for which NNLO corrections are very large,
and were thus excluded from the NNPDF4.0NLO dataset of
Ref.~\cite{NNPDF:2021njg}. However in this study we
prefer to compare PDFs produced using exactly the same code and the
same dataset, and that only differ in perturbative order and in the
presence of MHOUs, so that the effect of the latter can be assessed
without any confounding effect, however small.

Note also that the NNPDF4.0 NNLO without
MHOUs shown here are equivalent but not identical to the published NNPDF4.0
PDFs~\cite{NNPDF:2021njg}: they differ from them
because of the correction of a few minor bugs in the data
implementation, and because of the use of a new theory 
pipeline~\cite{Barontini:2023vmr} for the computation of
predictions, which in particular includes a new implementation of the
treatment of heavy quark mass effects that differs from the previous
one by subleading terms.
The impact of these changes was assessed in Appendix~A
of Ref.~\cite{NNPDF:2024djq}, and was found to be very limited, so that
for any application the NNPDF4.0 NNLO MHOU PDFs presented here can
be considered to be the counterpart of the published NNPDF4.0 NNLO
PDFs (without MHOU)~\cite{NNPDF:2021njg}.
  
\subsection{Fit quality}
\label{subsec:fit_quality}

In Tabs.~\ref{tab:chi2_TOTAL}-\ref{tab:chi2_OTHER} we report the number of
data points and the $\chi^2$ per data point in the NLO and NNLO NNPDF4.0
PDF determinations before and after inclusion of  MHOUs. When MHOUs are
not included, the covariance matrix is defined as in
Ref.~\cite{NNPDF:2021njg}, namely, it
is the sum of the experimental covariance matrix $C$ and of a theory
covariance matrix accounting for missing nuclear corrections $S^{\rm (nucl)}$, as
determined in Refs.~\cite{Ball:2018twp,Ball:2020xqw}, and whose impact
is discussed in
Section~8.6 of Ref.~\cite{NNPDF:2021njg}. When MHOUs are
included, the covariance matrix also contains the contribution
Eq.~(\ref{eq:thcovmat}) 
discussed in Sections~\ref{sec:thcovmat}-\ref{subsec:dataset}, that  we
call $S^{\rm (7pt)}$. 

Note that the MHOU  contribution is respectively excluded or included
both in the definition of the $\chi^2$ used by the NNPDF
algorithm (i.e.\ for pseudodata generation and in training and validation
loss functions),
and in the covariance matrix used in order to compute  the values
given in  Tabs.~\ref{tab:chi2_TOTAL}-\ref{tab:chi2_OTHER}. Note also
that the experimental covariance matrix
used in order to compute the values given in
Tabs.~\ref{tab:chi2_TOTAL}-\ref{tab:chi2_OTHER} differs from that used
in the NNPDF algorithm, because the
latter treats multiplicative uncertainties according to the $t_0$
method~\cite{Ball:2009qv} in order to avoid the d'Agostini bias, while
the former is just the published experimental covariance matrix.
In Tab.~\ref{tab:chi2_TOTAL} datasets are aggregated according to the
process categorization of Section~\ref{subsec:dataset}. Individual data sets
are displayed in Tab.~\ref{tab:chi2_DIS} (NC and CC DIS), in
Tab.~\ref{tab:chi2_DY} (NC and CC DY), and in Tab.~\ref{tab:chi2_OTHER}
(top pairs, single-inclusive jets, dijets,
isolated photons, and single top). The naming of the
datasets follows Ref.~\cite{NNPDF:2021njg}. Note finally that the
$\chi^2$ values shown in Tab.~\ref{tab:chi2_TOTAL}  cannot be obtained
by taking the weighted average of those from
Tabs.~\ref{tab:chi2_DIS}-\ref{tab:chi2_OTHER}, i.e. by adding each
$\chi^2$ value multiplied by the corresponding number of datapoints
and dividing the result by the total number of datapoints, because
several of the measurements reported in
Tabs.~\ref{tab:chi2_DIS}-\ref{tab:chi2_OTHER} are correlated to each
other, and furthermore, upon inclusion of the MHOUs, the covariance
matrix correlates all points to each other, as discussed in
Sect.~\ref{subsec:dataset} and shown in Fig.~\ref{fig:thcovmats}. These correlations are
lost when showing $\chi^2$ values for data subsets. For the same
reason, the total $\chi^2$ shown in  Tab.~\ref{tab:chi2_TOTAL} is not
the weighted average of individual values.

\begin{table}[!t]
  \scriptsize
  \centering
  \renewcommand{\arraystretch}{1.4}
  \begin{tabularx}{\textwidth}{Xrcccc}
  \toprule
    \multirow{2}{*}{Dataset}
  & \multirow{2}{*}{$N_{\rm dat}$ }
  & \multicolumn{2}{c}{NLO}
  & \multicolumn{2}{c}{NNLO}  \\
  & 
  & $C+S^{\rm (nucl)}$ & $C+S^{\rm (nucl)}+S^{\rm (7pt)}$
  & $C+S^{\rm (nucl)}$ & $C+S^{\rm (nucl)}+S^{\rm (7pt)}$ \\
  \midrule
  DIS NC
  & 2100
  & 1.30 & 1.22
  & 1.23 & 1.20 \\
  DIS CC
  &  989 
  & 0.92 & 0.87
  & 0.90 & 0.90 \\
  DY NC
  &  736 
  & 2.01 & 1.71
  & 1.20 & 1.15 \\
  DY CC
  &  157 
  & 1.48 & 1.42
  & 1.48 & 1.37 \\
  Top pairs
  &   64 
  & 2.08 & 1.24
  & 1.21 & 1.43 \\
  Single-inclusive jets
  &  356 
  & 0.84 & 0.82
  & 0.96 & 0.81 \\
  Dijets
  &  144 
  & 1.52 & 1.84
  & 2.04 & 1.71 \\
  Prompt photons 
  &   53 
  & 0.59 & 0.49
  & 0.75 & 0.67 \\
  Single top
  &   17 
  & 0.36 & 0.35
  & 0.36 & 0.38 \\
  \midrule
  Total
  & 4616 
  & 1.34 & 1.23
  & 1.17 & 1.13 \\
\bottomrule
\end{tabularx}

  \caption{The number of data points and the $\chi^2$ per data
    point for  the NLO and NNLO NNPDF4.0 PDF sets without and
    with MHOUs. Datasets are grouped according to
    the process categorization of Section~\ref{subsec:dataset}.}
  \label{tab:chi2_TOTAL}
\end{table}

\begin{table}[!t]
  \scriptsize
  \centering
  \renewcommand{\arraystretch}{1.4}
  \begin{tabularx}{\textwidth}{Xrcccc}
  \toprule
  \multirow{2}{*}{Dataset}
  & \multirow{2}{*}{$N_{\rm dat}$ }
  & \multicolumn{2}{c}{NLO}
  & \multicolumn{2}{c}{NNLO}  \\
  & 
  & $C+S^{\rm (nucl)}$ & $C+S^{\rm (nucl)}+S^{\rm (7pt)}$
  & $C+S^{\rm (nucl)}$ & $C+S^{\rm (nucl)}+S^{\rm (7pt)}$ \\
  \midrule
  NMC $F_2^d/F_2^p$
  &   121 
  & 0.87 & 0.87
  & 0.87 & 0.88 \\
  NMC $\sigma^{{\rm NC},p}$
  &   204 
  & 1.96 & 1.29
  & 1.62 & 1.33 \\
  SLAC $F_2^p$
  &     33
  & 1.72 & 0.84
  & 0.97 & 0.68 \\
  SLAC $F_2^d$
  &     34 
  & 1.08 & 0.75
  & 0.63 & 0.54 \\
  BCDMS $F_2^p$
  &  333 
  & 1.60 & 1.26
  & 1.41 & 1.29 \\
  BCDMS $F_2^d$
  &  248 
  & 1.06 & 1.01
  & 1.01 & 0.99 \\
  HERA I+II $\sigma_{\rm NC}\ e^-p$
  &  159 
  & 1.39 & 1.39
  & 1.39 & 1.39 \\
  HERA I+II $\sigma_{\rm NC}\ e^+p$ ($E_p=460$~GeV)
  &  204 
  & 1.11 & 1.04
  & 1.08 & 1.04 \\
  HERA I+II $\sigma_{\rm NC}\ e^+p$ ($E_p=575$~GeV)
  &  254 
  & 0.89 & 0.87
  & 0.92 & 0.88 \\
  HERA I+II $\sigma_{\rm NC}\ e^+p$ ($E_p=820$~GeV)
  &   70 
  & 1.08 & 0.96
  & 1.12 & 0.95 \\
  HERA I+II $\sigma_{\rm NC}\ e^+p$ ($E_p=920$~GeV)
  &  377 
  & 1.19 & 1.17
  & 1.30 & 1.25 \\
  HERA I+II $\sigma_{\rm NC}^{c}$
  &    37 
  & 1.83 & 1.66
  & 2.03 & 1.75 \\
  HERA I+II $\sigma_{\rm NC}^{b}$
  &   26 
  & 1.46 & 1.03
  & 1.45 & 1.11 \\
  \midrule
  CHORUS $\sigma_{CC}^{\nu}$
  &  416 
  & 0.96 & 0.95
  & 0.97 & 0.97 \\
  CHORUS $\sigma_{CC}^{\bar{\nu}}$
  &  416 
  & 0.90 & 0.87
  & 0.88 & 0.87 \\
  NuTeV $\sigma_{CC}^{\nu}$ (dimuon)
  &    39 
  & 0.22 & 0.22
  & 0.31 & 0.33 \\
  NuTeV $\sigma_{CC}^{\bar{\nu}}$ (dimuon)
  &    37 
  & 0.58 & 0.39
  & 0.56 & 0.64 \\
  HERA I+II $\sigma_{\rm CC}\ e^-p$
  &    42 
  & 1.39 & 1.18
  & 1.25 & 1.29 \\
  HERA I+II $\sigma_{\rm CC}\ e^+p$
  &    39 
  & 1.33 & 1.25
  & 1.22 & 1.25 \\
\bottomrule
\end{tabularx}

  \caption{Same as Tab.~\ref{tab:chi2_TOTAL}, for the DIS NC (top) and DIS CC
  (bottom) datasets.}
  \label{tab:chi2_DIS}
\end{table}

\begin{table}[!t]
  \scriptsize
  \centering
  \renewcommand{\arraystretch}{1.4}
  \begin{tabularx}{\textwidth}{Xrccrcc}
  \toprule
  \multirow{2}{*}{Dataset}
  & \multirow{2}{*}{$N_{\rm dat}$ }
  & \multicolumn{2}{c}{NLO}
  & \multicolumn{2}{c}{NNLO}  \\
  & 
  & $C+S^{\rm (nucl)}$ & $C+S^{\rm (nucl)}+S^{\rm (7pt)}$
  & $C+S^{\rm (nucl)}$ & $C+S^{\rm (nucl)}+S^{\rm (7pt)}$ \\
  \midrule
  E866 $\sigma^d/2\sigma^p$ (NuSea)
  & 15 
  & 0.66 & 0.52
  & 0.53 & 0.51 \\
  E866 $\sigma^p$ (NuSea)
  & 89 
  & 1.35 & 0.85
  & 1.63 & 1.00 \\
  E605 $\sigma^d/2\sigma^p$ (NuSea)
  & 85 
  & 0.44 & 0.42
  & 0.46 & 0.45 \\
  E906 $\sigma^d/2\sigma^p$ (SeaQuest)
  & 6 
  & 1.23 & 3.20
  & 0.90 & 0.90 \\
  CDF $Z$ differential
  & 28 
  & 1.36 & 1.26
  & 1.23 & 1.18 \\
  D0 $Z$ differential
  & 28 
  & 0.75 & 0.73
  & 0.64 & 0.64 \\
  ATLAS low-mass DY 7 TeV
  & 6 
  & 13.3 & 8.97
  & 0.87 & 0.78 \\
  ATLAS high-mass DY 7 TeV
  & 5 
  & 1.60 & 1.64
  & 1.60 & 1.67 \\
  ATLAS $Z$ 7 TeV ($\mathcal{L}=35$~pb$^{-1}$)
  & 8 
  & 0.77 & 0.49
  & 0.60 & 0.57 \\
  ATLAS $Z$ 7 TeV ($\mathcal{L}=4.6$~fb$^{-1}$) CC
  & 24 
  & 5.00 & 3.29
  & 1.73 & 1.68 \\
  ATLAS $W,Z$ 7 TeV ($\mathcal{L}=4.6$~fb$^{-1}$) CF
  & 15 
  & 1.82 & 1.21
  & 1.07 & 1.02 \\
  ATLAS low-mass DY 2D 8 TeV
  & 60 
  & 1.73 & 1.04
  & 1.21 & 1.08 \\
  ATLAS high-mass DY 2D 8 TeV
  & 48 
  & 1.48 & 1.34
  & 1.12 & 1.08 \\
  ATLAS $\sigma_{Z}^{\rm tot}$ 13 TeV
  & 1 
  & 0.48 & 0.43
  & 0.30 & 0.60 \\
  ATLAS $Z$ $p_T$ 8 TeV ($p_T,m_{\ell\ell}$)
  & 44 
  & 1.05 & 0.93
  & 0.90 & 0.91 \\
  ATLAS $Z$ $p_T$ 8 TeV ($p_T,y_Z$)
  & 48 
  & 0.74 & 0.69
  & 0.88 & 0.70 \\
  CMS DY 2D 7 TeV
  & 110 
  & 3.66 & 1.10
  & 1.35 & 1.32 \\
  CMS $Z$ $p_T$ 8 TeV
  & 28 
  & 1.66 & 1.58
  & 1.40 & 1.41 \\
  LHCb $Z\to ee$ 7 TeV
  & 9 
  & 1.51 & 1.36
  & 1.64 & 1.53 \\
  LHCb $Z \to \mu$ 7 TeV
  & 15 
  & 1.01 & 0.85
  & 0.78 & 0.73 \\
  LHCb $Z\to ee$ 8 TeV
  & 17 
  & 1.67 & 1.21
  & 1.25 & 1.26 \\
  LHCb $Z\to \mu$ 8 TeV
  & 16 
  & 1.40 & 1.05
  & 1.46 & 1.59 \\
  LHCb $Z\to ee$ 13 TeV
  & 16 
  & 1.34 & 1.61
  & 0.96 & 1.80 \\
  LHCb $Z\to \mu\mu$ 13 TeV
  & 15 
  & 1.88 & 1.13
  & 1.75 & 0.99 \\
  \midrule
  D0 $W$ muon asymmetry
  & 9 
  & 2.48 & 1.92
  & 1.99 & 1.95 \\
  ATLAS $W$ 7 TeV ($\mathcal{L}=35$~pb$^{-1}$)
  & 22 
  & 1.23 & 1.15
  & 1.13 & 1.12 \\
  ATLAS $W$ 7 TeV ($\mathcal{L}=4.6$~fb$^{-1}$)
  & 22 
  & 2.74 & 2.26
  & 2.15 & 2.16 \\
  ATLAS $\sigma_{W}^{\rm tot}$ 13 TeV
  & 2 
  & 0.10 & 0.40
  & 1.21 & 1.60 \\
  ATLAS $W^+$+jet 8 TeV
  & 15 
  & 1.68 & 1.15
  & 0.79 & 0.79 \\
  ATLAS $W^-$+jet 8 TeV
  & 15 
  & 1.82 & 1.31
  & 1.49 & 1.45 \\
  CMS $W$ electron asymmetry 7 TeV
  & 11 
  & 0.85 & 0.96
  & 0.83 & 0.85 \\
  CMS $W$ muon asymmetry 7 TeV
  & 11 
  & 2.05 & 1.75
  & 1.74 & 1.73 \\
  CMS $W$ rapidity 8 TeV
  & 22 
  & 0.92 & 0.71
  & 1.39 & 1.03 \\
  LHCb $W \to \mu$ 7 TeV
  & 14 
  & 1.76 & 1.44
  & 2.76 & 1.99 \\
  LHCb $W \to \mu$ 8 TeV
  & 14 
  & 0.76 & 0.51
  & 0.96 & 0.92 \\
\bottomrule
\end{tabularx}

  \caption{Same as Tab.~\ref{tab:chi2_TOTAL}, for the DY NC (top) and DY CC
  (bottom) datasets.}
  \label{tab:chi2_DY}
\end{table}

\begin{table}[!t]
  \scriptsize
  \centering
  \renewcommand{\arraystretch}{1.4}
  \begin{tabularx}{\textwidth}{Xrccrcc}
  \toprule
  \multirow{2}{*}{Dataset}
  & \multirow{2}{*}{$N_{\rm dat}$ }
  & \multicolumn{2}{c}{NLO}
  & \multicolumn{2}{c}{NNLO}  \\
  & 
  & $C+S^{\rm (nucl)}$ & $C+S^{\rm (nucl)}+S^{\rm (7pt)}$
  & $C+S^{\rm (nucl)}$ & $C+S^{\rm (nucl)}+S^{\rm (7pt)}$ \\
  \midrule
  ATLAS $\sigma_{tt}^{\rm tot}$ 7 TeV
  & 1 
  & 11.7 & 3.66
  & 4.66 & 2.40 \\
  ATLAS $\sigma_{tt}^{\rm tot}$ 8 TeV
  & 1 
  & 2.28 & 0.87
  & 0.03 & 0.03 \\
  ATLAS $\sigma_{tt}^{\rm tot}$ 13 TeV ($\mathcal{L}$=139~fb$^{-1}$)
  & 1 
  & 4.58 & 1.18
  & 0.56 & 0.41 \\
  ATLAS $t\bar{t}~\ell$+jets 8 TeV ($1/\sigma d\sigma/dy_t$)
  & 4 
  & 3.39 & 1.89
  & 3.01 & 3.70 \\
  ATLAS $t\bar{t}~\ell$+jets 8 TeV ($1/\sigma d\sigma/dy_{t\bar t}$)
  & 4 
  & 7.19 & 3.85
  & 3.65 & 5.80 \\
  ATLAS $t\bar{t}~2\ell$ 8 TeV ($1/\sigma d\sigma/dy_{t\bar t}$)
  & 4 
  & 1.80 & 1.76
  & 1.57 & 1.86 \\
  CMS $\sigma_{tt}^{\rm tot}$ 5 TeV
  & 1 
  & 0.72 & 0.95
  & 0.01 & 0.01 \\
  CMS $\sigma_{tt}^{\rm tot}$ 7 TeV
  & 1 
  & 6.37 & 1.82
  & 1.10 & 0.50 \\
  CMS $\sigma_{tt}^{\rm tot}$ 8 TeV
  & 1 
  & 4.39 & 1.21
  & 0.31 & 0.17 \\
  CMS $\sigma_{tt}^{\rm tot}$ 13 TeV
  & 1 
  & 1.06 & 0.36
  & 0.04 & 0.01 \\
  CMS $t\bar{t}~\ell$+jets 8 TeV ($1/\sigma d\sigma/dy_{t\bar t}$)
  & 9 
  & 1.67 & 1.61
  & 1.20 & 1.59 \\
  CMS $t\bar{t}$ 2D $2\ell$ 8 TeV ($1/\sigma d\sigma/dy_tdm_{t\bar t}$)
  & 15 
  & 2.03 & 1.84
  & 1.32 & 1.25 \\
  CMS $t\bar{t}~2\ell$ 13 TeV ($d\sigma/dy_t$)
  & 10 
  & 0.77 & 0.71
  & 0.51 & 0.59 \\
  CMS $t\bar{t}~\ell$+jet 13 TeV ($d\sigma/dy_t$)
  & 11 
  & 0.54 & 0.26
  & 0.56 & 0.66 \\ 
  \midrule
  ATLAS incl. jets 8~TeV, $R=0.6$
  & 171 
  & 0.70 & 0.73
  & 0.71 & 0.64 \\
  CMS incl. jets 8 TeV
  & 185 
  & 0.97 & 0.81
  & 1.19 & 0.95 \\
  \midrule
  ATLAS dijets 7 TeV, $R=0.6$
  & 90 
  & 1.48 & 1.82
  & 2.16 & 1.69 \\
  CMS dijets 7 TeV
  & 54 
  & 1.59 & 2.07
  & 1.84 & 1.74 \\
  \midrule
  ATLAS isolated $\gamma$ prod. 13 TeV
  & 53 
  & 0.59 & 0.50
  & 0.75 & 0.67 \\
  \midrule
  ATLAS single~$t$ $R_{t}$ 7 TeV
  & 1 
  & 0.38 & 0.30
  & 0.48 & 0.57 \\
  ATLAS single~$t$ $R_{t}$ 13 TeV
  & 1 
  & 0.05 & 0.03
  & 0.06 & 0.07 \\
  ATLAS single~$t$ 7 TeV ($1/\sigma d\sigma/dy_t$)
  & 3 
  & 0.84 & 0.82
  & 0.97 & 0.94 \\
  ATLAS single~$t$ 7 TeV ($1/\sigma d\sigma/dy_{\bar t}$)
  & 3 
  & 0.06 & 0.06
  & 0.06 & 0.06 \\
  ATLAS single~$t$ 8 TeV ($1/\sigma d\sigma/dy_t$)
  & 3 
  & 0.35 & 0.33
  & 0.24 & 0.26 \\
  ATLAS single~$t$ 8 TeV ($1/\sigma d\sigma/dy_{\bar t}$)
  & 3 
  & 0.19 & 0.21
  & 0.19 & 0.19 \\
  CMS single~$t$ $\sigma_{t}+\sigma_{\bar{t}}$ 7 TeV
  & 1 
  & 0.91 & 0.96
  & 0.76 & 0.84 \\
  CMS single~$t$ $R_{t}$ 8 TeV
  & 1 
  & 0.13 & 0.09
  & 0.17 & 0.20 \\
  CMS single~$t$ $R_{t}$ 13 TeV
  & 1 
  & 0.32 & 0.28
  & 0.35 & 0.38 \\
\bottomrule
\end{tabularx}

  \caption{Same as Tab.~\ref{tab:chi2_TOTAL}, for (from top to
    bottom)
    top pair,
    single-inclusive jet, isolated photon
     and single top production.}
  \label{tab:chi2_OTHER}
\end{table}

Tables~\ref{tab:chi2_TOTAL}-\ref{tab:chi2_OTHER} show that upon
inclusion of the MHOU  covariance matrix  the total
$\chi^2$ decreases for both the NLO and NNLO fits, but  the decrease is
more substantial at NLO. Even after inclusion of the MHOU, the
NLO $\chi^2$ remains somewhat higher than the NNLO one. Inspection of
Tab.~\ref{tab:chi2_DIS}-\ref{tab:chi2_OTHER}  shows that this is in
fact due to a small number of datasets (specifically ATLAS low-mass
Drell-Yan), and we have further verified that this is in turn
due to a small number
of very accurately measured data points (excluded by the NLO cuts of
Ref.~\cite{NNPDF:2021njg}) for which NNLO corrections are very
substantially underestimated  by scale variation.  However, for the majority
of datapoints and of
process categories,  the MHOU covariance matrix correctly accounts
for the mismatch 
between data and theory predictions at NLO due to missing
NNLO  terms, consistently with the validation of
Section~\ref{sec:validation}. 

\subsection{PDFs and PDF uncertainties}
\label{subsec:perturbative_convergence}

\begin{figure}[!p]
  \centering
  \includegraphics[width=0.45\textwidth]{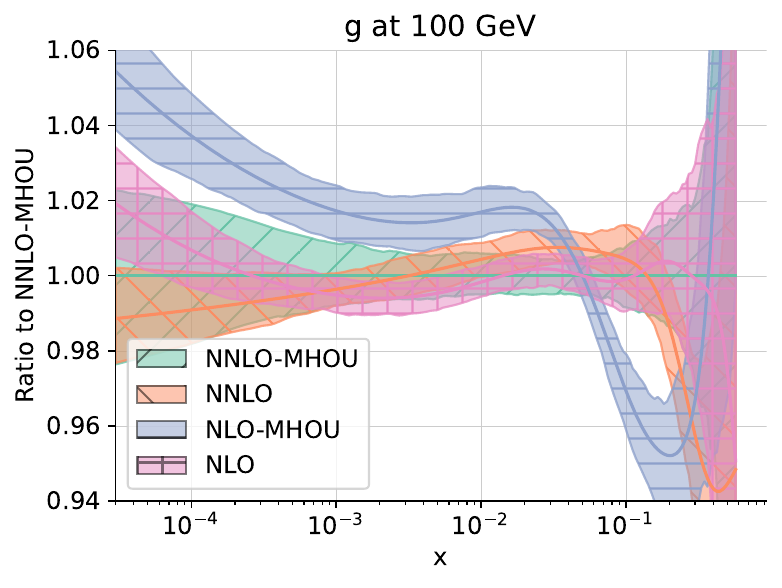}
  \includegraphics[width=0.45\textwidth]{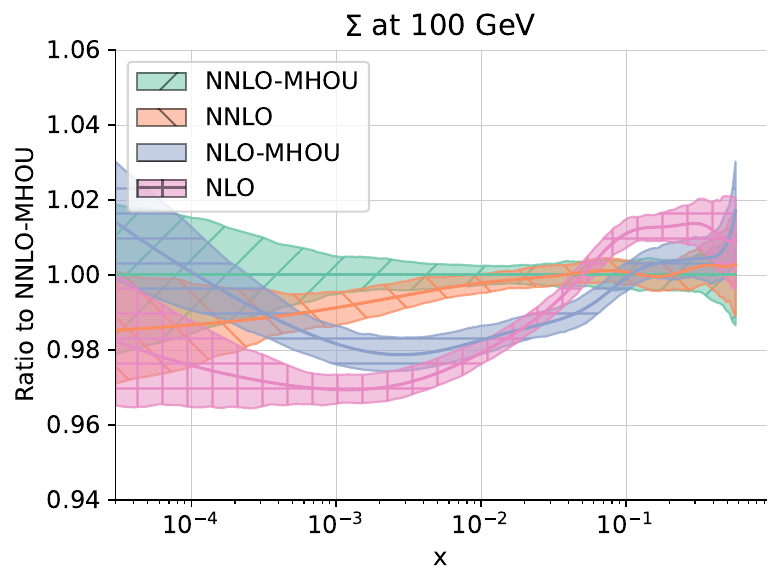}
  \includegraphics[width=0.45\textwidth]{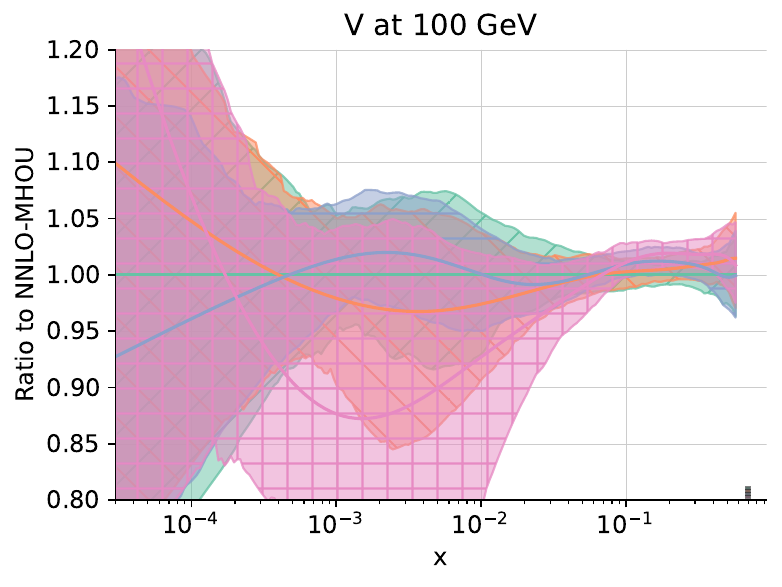}
  \includegraphics[width=0.45\textwidth]{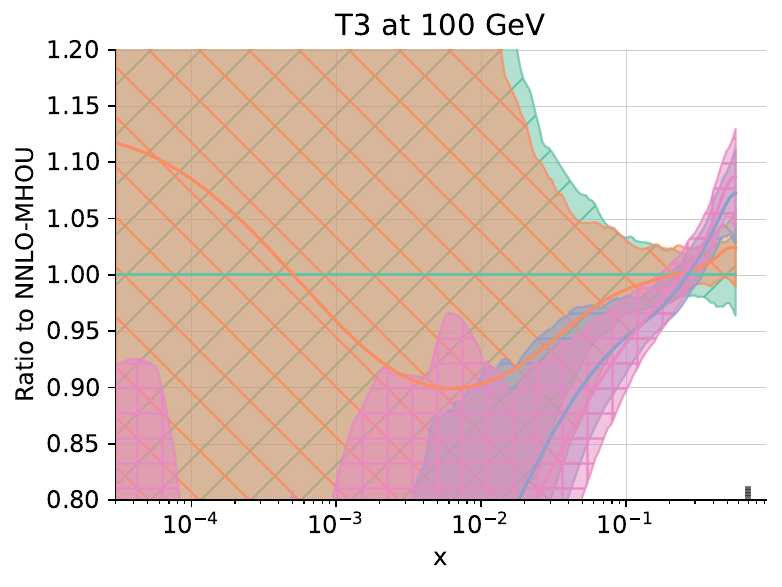}
    \includegraphics[width=0.45\textwidth]{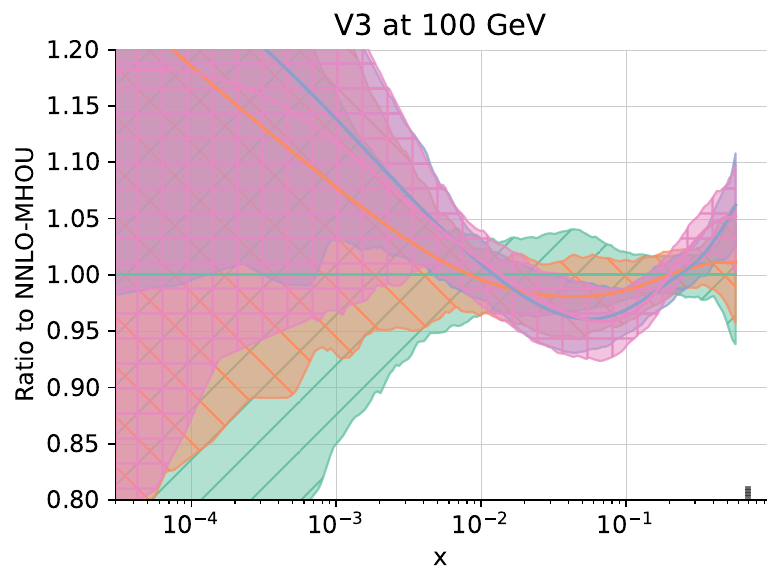}
    \includegraphics[width=0.45\textwidth]{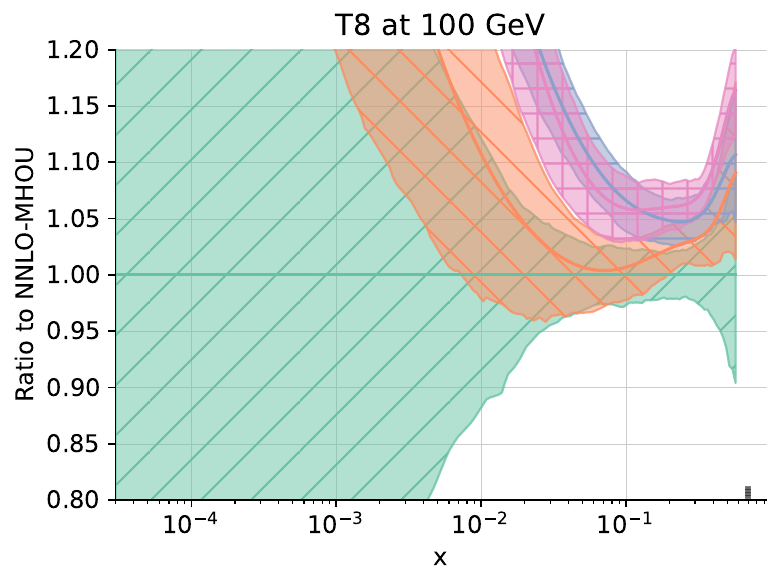}
      \includegraphics[width=0.45\textwidth]{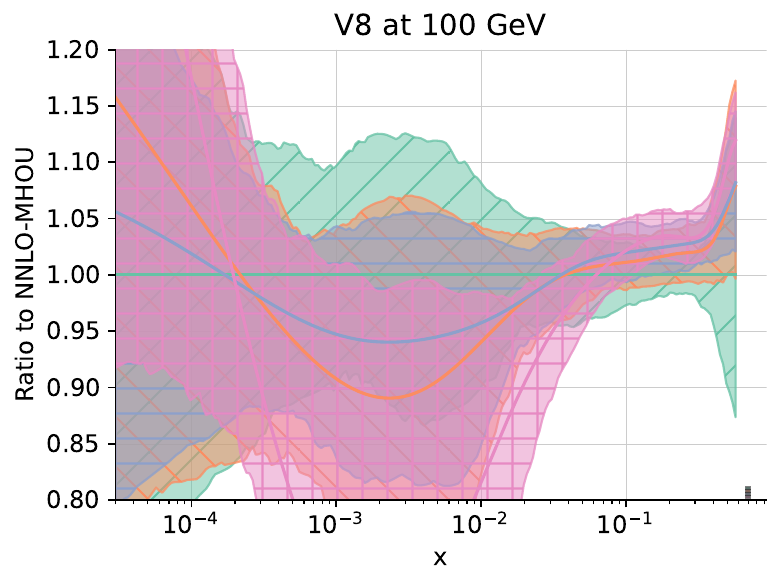}
  \includegraphics[width=0.45\textwidth]{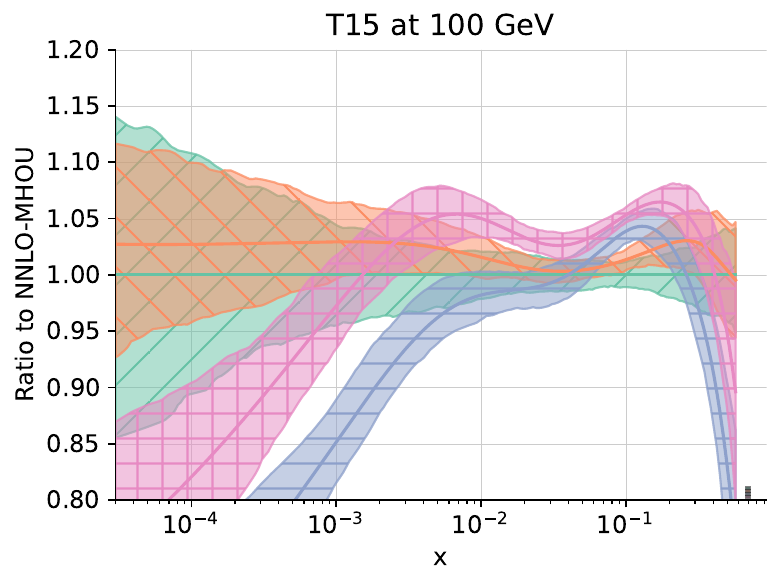}
  \caption{The NLO and NNLO PDFs  with and without MHOUs at $Q=100$~GeV
    determined in this work. The gluon, singlet, valence
    ($V$, $V_3$, $V_8$), and triplet ($T_3$, $T_8$, $T_{15}$) PDFs are
    shown. All curves are  normalized to the NNLO with MHOUs. The
    bands correspond to one sigma uncertainty.}
  \label{fig:pdfs}
\end{figure}

\begin{figure}[!p]
  \centering
  \includegraphics[width=0.45\textwidth]{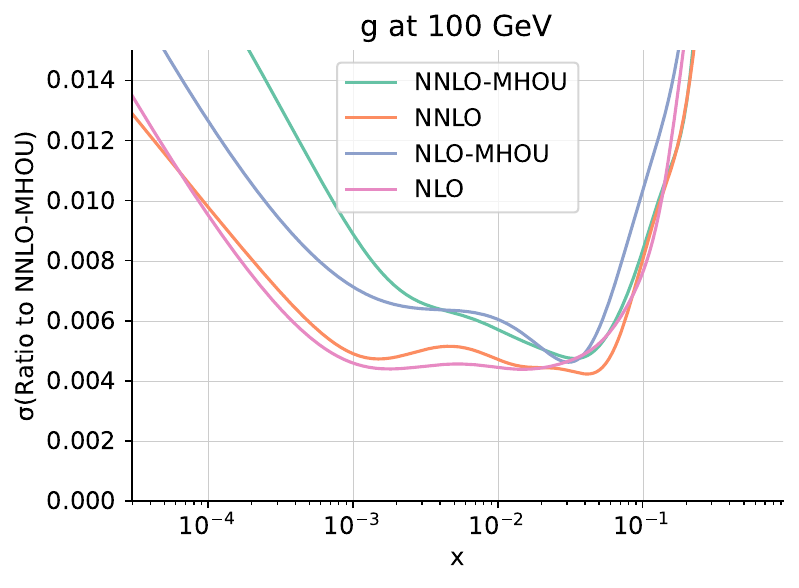}
  \includegraphics[width=0.45\textwidth]{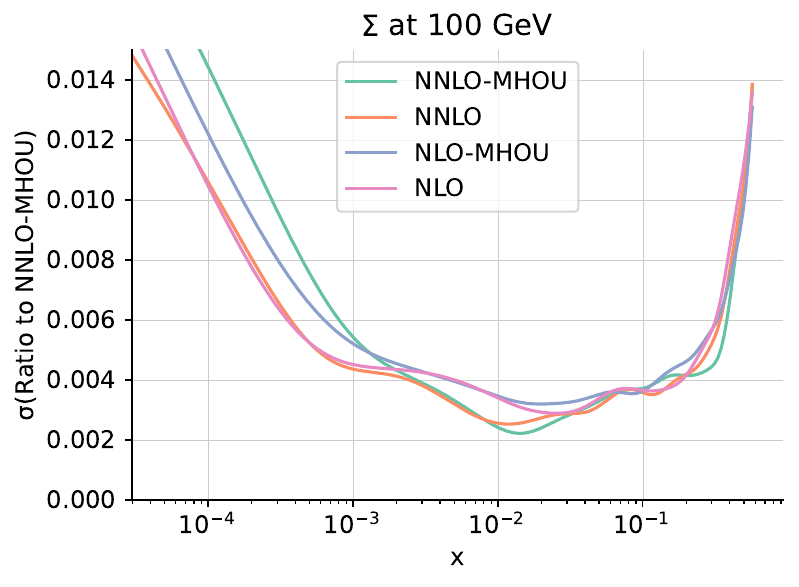}\\
  \includegraphics[width=0.45\textwidth]{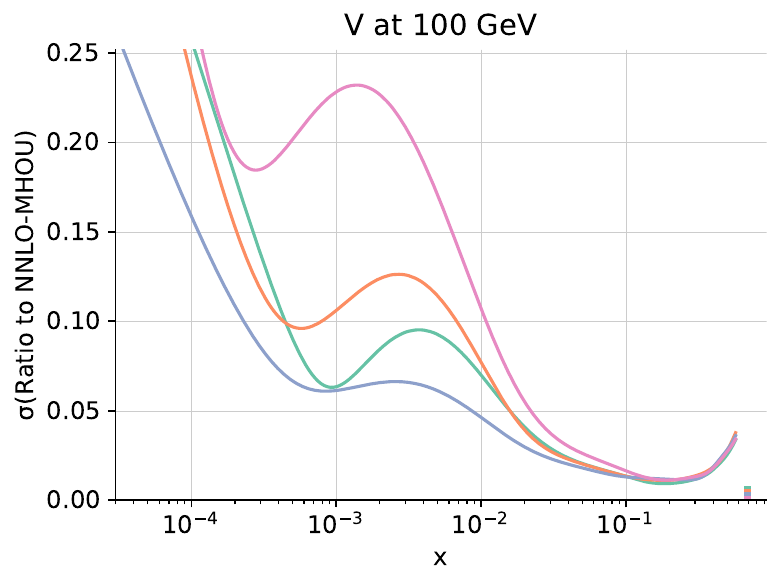}
  \includegraphics[width=0.45\textwidth]{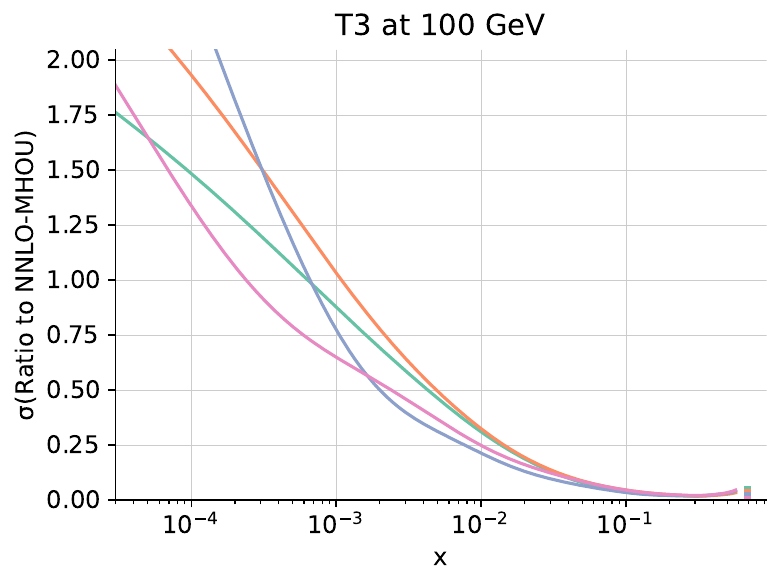}\\
  \includegraphics[width=0.45\textwidth]{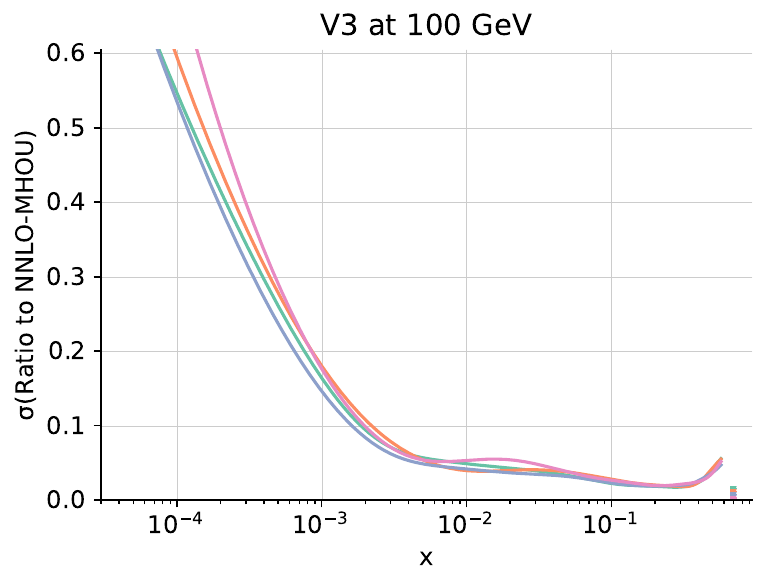}
  \includegraphics[width=0.45\textwidth]{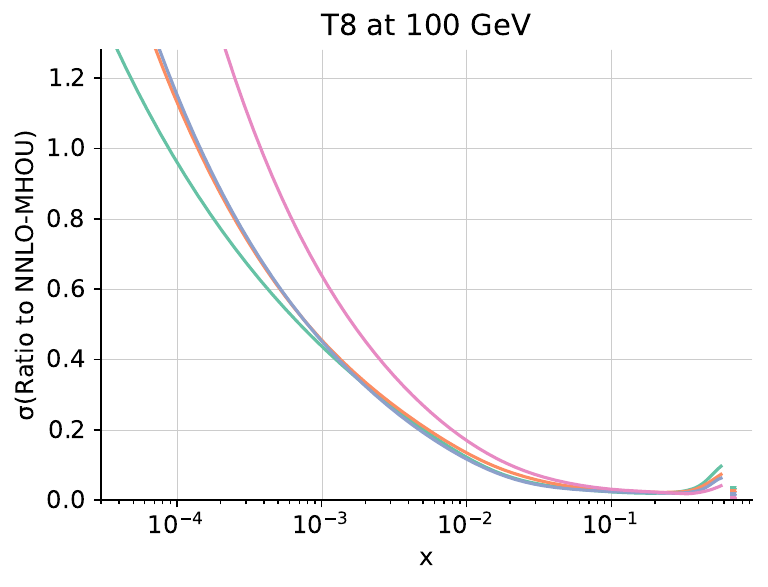}\\
  \includegraphics[width=0.45\textwidth]{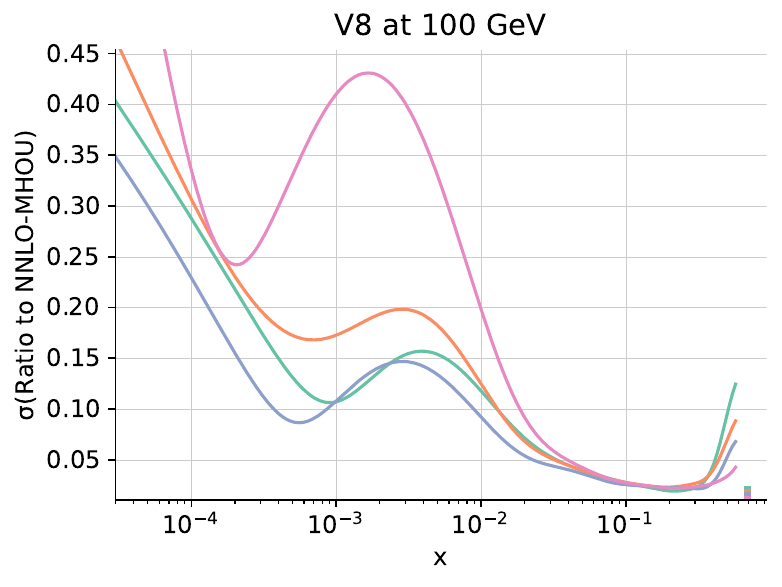}
  \includegraphics[width=0.45\textwidth]{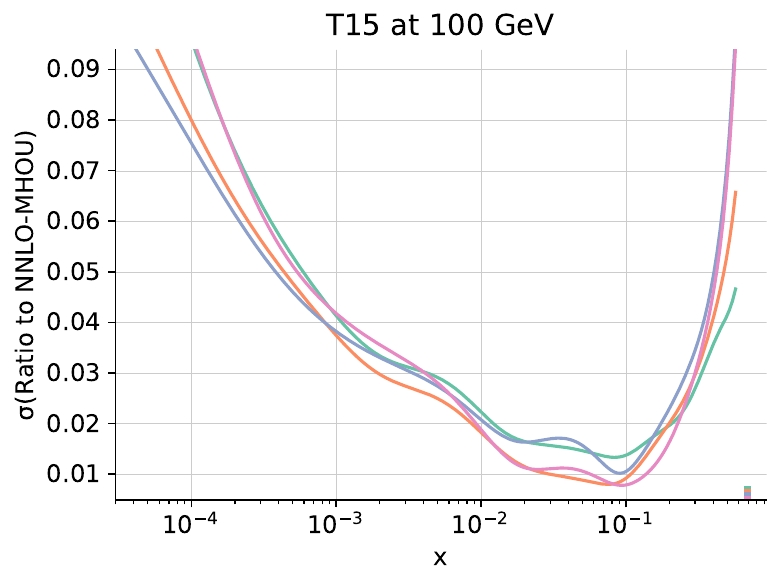}\\
  \caption{Relative one sigma uncertainties for the PDFs shown in
    Fig.~\ref{fig:pdfs}. All uncertainties are  normalized to the corresponding
    central
    NNLO PDFs with MHOUs.}
  \label{fig:pdf_uncertainties}
\end{figure}

Individual PDFs at NLO and NNLO, with and without MHOUs, are compared in
Fig.~\ref{fig:pdfs} at $Q=100$~GeV. We
show the gluon, singlet, valence ($V$, $V_3$, $V_8$), and triplet ($T_3$,
$T_8$, $T_{15}$) distributions (see Section~3.1.1 of
Ref.~\cite{NNPDF:2021njg}), all shown as a ratio to the  NNLO PDFs
with MHOUs. The corresponding one sigma uncertainties are shown in
Fig.~\ref{fig:pdf_uncertainties}.
The change in central value due to the inclusion of MHOUs
is  generally moderate at NNLO; at  NLO it is significant for the
gluon and singlet, but quite moderate for all other PDF
combinations.

Inspection of Fig.~\ref{fig:pdf_uncertainties} shows that the PDF
uncertainty at NNLO in the data region remains on average unchanged 
upon inclusion of MHOUs, though in the singlet
sector it increases at small $x$, especially for the gluon where the
increase is up to $x\sim 0.1$. At NLO the uncertainty is
generally reduced in the nonsinglet sector, while in the singlet sector the
uncertainty increases for all $x$, especially for the gluon.
This is 
consistent with the observation of Sect.~\ref{subsec:fit_quality}
that at NLO the MHOU from scale variation does not fully account for
the large shift from NLO to NNLO for some datasets. The somewhat
counter-intuitive fact that the
uncertainty on the PDF does not increase and may even  be reduced upon
inclusion of an extra source of 
uncertainty in the $\chi^2$ was already observed in
Refs.~\cite{Ball:2018twp,Ball:2020xqw} and demonstrates the increased
compatibility of the data due to the MHOU.

\begin{table}[!t]
  \scriptsize
  \centering
  \renewcommand{\arraystretch}{1.4}
  \begin{tabularx}{\textwidth}{Xcccc}
  \toprule
  \multirow{2}{*}{Dataset}
  & \multicolumn{2}{c}{NLO}
  & \multicolumn{2}{c}{NNLO}  \\
  & $C+S^{\rm (nucl)}$ & $C+S^{\rm (nucl)}+S^{\rm (7pt)}$
  & $C+S^{\rm (nucl)}$ & $C+S^{\rm (nucl)}+S^{\rm (7pt)}$ \\
  \midrule
  DIS NC
  & 0.14 & 0.13 
  & 0.15 & 0.13 \\
  DIS CC
  & 0.12 & 0.12 
  & 0.12 & 0.12 \\
  DY NC
  & 0.19 & 0.17 
  & 0.18 & 0.17 \\
  DY CC
  & 0.37 & 0.30 
  & 0.35 & 0.32 \\
  Top pairs
  & 0.19 & 0.16 
  & 0.17 & 0.17 \\
  Single-inclusive jets
  & 0.13 & 0.12 
  & 0.13 & 0.13 \\
  Dijets
  & 0.10 & 0.09 
  & 0.11 & 0.10 \\
  Prompt photon 
  & 0.06 & 0.06 
  & 0.06 & 0.06 \\
  Single top
  & 0.04 & 0.04 
  & 0.04 & 0.04 \\
  \midrule
  Total
  & 0.16 & 0.15
  & 0.17 & 0.15 \\
\bottomrule
\end{tabularx}

  \caption{The $\phi$ estimator Eq.~(\ref{eq:phidef}) for PDFs at
    NLO and NNLO with and without
    MHOUs for the process categories of 
    Section~\ref{subsec:dataset}.}
  \label{tab:phi_TOTAL}
\end{table}

The effect of the inclusion of MHOUs on PDF uncertainties is
both $x$- and PDF-dependent, hence in order to obtain an overall
quantitative assessment it is necessary to look at the PDF uncertainty
on physics predictions. This 
can be obtained through the 
$\phi$ estimator, which was introduced in
Ref.~\cite{NNPDF:2014otw}, and is defined as
\begin{equation}
\phi_{\chi^2} =\sqrt{\langle \chi^2\rangle - \chi^2},
\label{eq:phidef}
\end{equation}
where by $\langle \chi^2\rangle $ we denote the average value of the
$\chi^2$ (per datapoint) evaluated for each single replica and
averaged over replicas, while $\chi^2$ is the value shown in
Tab.~\ref{tab:chi2_TOTAL} and computed using   the
best-fit PDF, i.e. the average over replicas. For a single datapoint, $\phi$ is just the
ratio of the PDF uncertainty over the data uncertainty; for many
uncorrelated datapoints it is the square root of the
average value of the ratio of the PDF variance to the data variance; and
for correlated datapoints it is this quantity when computed in the
basis of eigenvectors of the experimental covariance matrix, thus averaging
ratios of the diagonal elements  of the theory covariance matrix in this basis
to the eigenvalues of the experimental covariance matrix.  
Hence, $\phi$ directly measures the PDF uncertainty on the
predictions in units of the experimental uncertainties, and thus
provides an estimate of the consistency of the data. Indeed, a
value $\phi<1$ means that on average the uncertainties in the predictions
are  smaller than those of the original data, indicating that 
consistent data are being combined successfully by the underlying theory
(see also the discussion in Section~6 of
Ref.~\cite{NNPDF:2019ubu}).

The value of $\phi$ before and after inclusion of the
MHOUs is shown at NLO and NNLO in
Tab.~\ref{tab:phi_TOTAL}. It is clear that
upon inclusion of MHOUs $\phi$  is always either unchanged or
reduced. The reduction is more significant for processes that are
sensitive to nonsinglet combinations, such as charged-current
Drell-Yan, in agreement with the behavior of the PDF uncertainties of
Fig.~\ref{fig:pdf_uncertainties}, and on average it is more
marked at NNLO than at NLO.

The reduction of  $\phi$  means that PDF uncertainties on physical
predictions in the
data region are reduced on average by the inclusion of MHOUs. It is
interesting to observe that this reduction, while apparent at NLO at
least for some PDF combination, is not always visible at NNLO in the
PDF plots of 
Figs.~\ref{fig:pdfs}-\ref{fig:pdf_uncertainties}, where instead an
increased uncertainty is seen, especially in the singlet sector.
It should be however
observed that the the uncertainty displayed in 
Figs.~\ref{fig:pdfs}-\ref{fig:pdf_uncertainties} is the diagonal
uncertainty, which combines correlated and uncorrelated uncertainties
in quadrature. On the other hand, the computation of physical
observables involves 
different PDF combinations, and beyond LO always an integration over
different values of the
momentum fraction, all of which are correlated to each other. These
correlations are fully included in the $\phi$ indicator.
Now, as  seen in Fig.~\ref{fig:thcovmats}, MHOUs are
generally highly correlated: for example, because MHOUs are smooth,
they have a similar impact on datapoints which are kinematically
close. Of course, the correlated uncertainty is always smaller or
equal to the uncorrelated one. Hence the different behavior of
diagonal PDF uncertainties and $\phi$ shows that the correlation of
MHOUs leads in
turn to highly correlated MHOUs on PDFs. Quite apart from this, note that
of course the $\phi$ indicator does not
provide any information on the behavior of uncertainties in the
extrapolation region, hence when computing physical predictions
outside the kinematic region covered by the current dataset the PDF
uncertainty may well increase upon inclusion of MHOUs.

It is interesting to contrast the behavior of the $\phi$ indicator
seen in Tab.~\ref{tab:phi_TOTAL} to that which  was discussed
in Ref.~\cite{NNPDF:2019ubu} (at NLO only), see
 Sect.~6 (Tab.~6 and especially Tab.~8) of that work. Firstly, it should be noticed that the 
$\phi$ value before inclusion of MHOU in that reference was more than
twice as large as it is here. This is due to the fact that
uncertainties in the NNPDF4.0 PDF set, discussed here, are
significantly smaller than in the then available NNPDF3.1 PDF set. The
reason is the
improvement in methodology, 
even with fixed underlying dataset, as extensively discussed in
Sect.~8.2  (see Fig.~46) of Ref.~\cite {NNPDF:2021njg}. Indeed, for
NNPDF4.0 NNLO, $\phi=0.16$ (Tab.~31 of Ref.~\cite{NNPDF:2021njg})
while for NNPDF3.1 NNLO, $\phi=0.36$ (Tab.~8 of Ref.~\cite{NNPDF:2019ubu}).

Furthermore, in Ref.~\cite{NNPDF:2019ubu} it was observed that  
upon addition of a MHOU term to the covariance matrix used in the fit
one would expect the uncertainty of the result to increase by an
amount which for a single datapoint would just be the sum in
quadrature of the MHOU term and the experimental uncertainty. For
several correlated measurements, this can
again be formalized in terms of an expected increased of the $\phi$
indicator based on the experimental and MHOU covariance matrices (see
Eq.~(6.5) of Ref.~\cite{NNPDF:2019ubu}). The value of $\phi$ was then
observed to increase upon the addition of MHOUs, but by less than the
expected amount, and this was interpreted as a sign of the increased
compatibility of the data upon inclusion of the MHOUs.

Here instead upon inclusion of MHOUs the value of $\phi$ is actually
reduced,  and this despite the fact
that with NNPDF4.0 methodology the value of  $\phi$ is lower to begin
with. This suggests that the improvement in data compatibility is now
rather more significant. This is surely at least in part  due to the
fact that whereas for NNPDF3.1 hadron collider data played a 
subdominant role in comparison to DIS data, the converse is true for
NNPDF4.0 (see the discussion in Sects.~7.2.4 and 7.2.5. of
Ref.~\cite{NNPDF:2021njg}). Because higher-order corrections are more
substantial for hadronic processes than for DIS, the impact of MHOUs is
accordingly enhanced. Also, because uncertainties with NNPDF4.0
methodology are smaller for equal data uncertainties, the effect of
tension between data from different processes due to MHO corrections
is enhanced, and the impact of the improved compatibility upon
inclusion of MHOUs accordingly enhanced.
We conclude that the evidence points towards the fact that
data compatibility is increased by the
inclusion of PDF uncertainties, especially at NNLO.

A priori, PDF sets with and
without MHOUs should not necessarily
be compatible within uncertainties, given that
the latter do not include an existing source of uncertainty. As a matter of
fact, they  do agree in the nonsinglet sector, where  MHOU are at most
of about the same 
size as the PDF uncertainty before inclusion of MHOUs, but they
generally do not in the singlet sector, where NNLO corrections can be
very large.

Inclusion of MHOUs generally moves the NLO PDFs towards the NNLO,
thereby improving perturbative convergence, except for the  gluon.
In fact, even for the singlet, while the NLO moves towards the NNLO
upon inclusion of 
MHOUs, the NNLO result remains well outside the NLO uncertainty band,
especially  at small $x$. Again, this shows that in the singlet sector
there are large NNLO corrections to the NLO result that are
underestimated by MHOUs determined through scale variation. At small
$x$ this can be understood as the consequence of unresummed small-$x$
logarithms~\cite{Ball:2017otu} whose increase with perturbative order
is  not accounted for by scale variation.

The general conclusion is thus that the inclusion of MHOUs estimated
through scale variation improves
data compatibility. This results in a
moderate shift of PDF central values, and a   reduction of PDF uncertainties
on physics predictions 
in the data region, demonstrated by a reduction of the $\phi$
indicator,  with diagonal PDF uncertainties mostly unchanged at NNLO
and reduced in the nonsinglet sector at NLO, but generally somewhat
increased for the gluon both at NLO and NNLO.
At NLO the inclusion of MHOUs manages to account for the
effect of MHO terms on fit quality while having a moderate impact on
PDF uncertainties and central values in the nonsinglet sector,
and a more significant impact on central values with an increase in
uncertainties in the singlet sector, while not fully accounting for the
largest missing NNLO corrections.  In the small-$x$ extrapolation
region PDF uncertainties generally increase upon inclusion of MHOUs, both
at NLO and NNLO.

A more detailed study of perturbative stability and the effect of the
inclusion of MHOUs on it is reserved to a companion
publication~\cite{NNPDF:2024nan}, in which the PDF determination and
consequently the results presented here are extended to N$^3$LO. We
also refer to this work for more extensive PDF comparisons, 
including comparisons at the level of flavor-basis PDFs and parton
luminosities, as well as for first studies of the implication of the
inclusion of MHOUs at various perturbative orders on  predictions for
LHC cross-sections.

\section{Delivery and outlook}
\label{sec:summary}

We have presented NLO and NNLO sets of parton distributions
for which the PDF uncertainty includes not only the uncertainty coming
from the data and that from the analysis methodology used to
go from the data to the PDFs, but also the uncertainty coming from the
perturbative truncation of the computations used in order to get the 
theory predictions that are compared to data (MHOUs). We followed the
methodology that was developed in
Refs.~\cite{NNPDF:2019vjt,NNPDF:2019ubu} and used there to construct the
first NLO PDF sets including MHOUs. This methodology can now be used to
determine MHOUs up to N$^3$LO, thanks to the
availability of the EKO evolution code~\cite{Candido:2022tld}, and the
interfacing of the NNPDF code~\cite{NNPDF:2021uiq} both to it and to flexible
tools for the computation of
physical processes (such as the YADISM module~\cite{yadism} for DIS) through
a new and streamlined theory pipeline~\cite{Barontini:2023vmr}.

The MHOUs on PDFs discussed in this paper
should be  treated as an extra contribution
to the PDF uncertainty: they reflect the uncertainty in the theory
predictions used in PDF determination and are thus on a par with the
experimental uncertainty on the data themselves. They are consequently
independent of the further MHOU on the hard cross-section of the
processes which are being predicted. 
The total uncertainty on predictions must therefore be
obtained by combining the
PDF uncertainty, now also including a MHOU component,  with the
MHOU uncertainty on the hard cross section computation. The latter is
typically
determined as the
envelope of a 7-point scale variation (see
e.g.\ Ref.~\cite{deFlorian:2016spz}). However, it is also possible to include
MHOUs on the hard cross section by constructing a theory covariance
matrix for the hard corss section itself. An advantage of doing so is
that it is then also possible to
keep into account  the correlation between the theory uncertainty in
the process used for PDF determination, and that on the
hard cross-section, which might
become relevant if 
the experimental uncertainties are small and the predicted process was
also used for PDF
determination~\cite{Harland-Lang:2018bxd,Kassabov:2022orn}.  This can
be done by 
determining the cross-correlation of MHO and PDF uncertainties
 between the predicted process and those
 used for PDF determination~\cite{Pearson:2021lmm}.

 In this respect,
 it is interesting to observe that an altogether different option for
 the inclusion of MHOUs on predictions that accounts for MHOUs on
 PDFs, and their full correlation to MHOUs   on the
 hard cross-sections, is to include the scale
 variation in the Monte Carlo sampling~\cite{Kassabov:2022orn}. A
 comparative study of this methodology to that adopted in the present
 paper, as well of different prescriptions for scale variation (such
 as, for instance different prescrriptions for the connstruction of
 the theory covariance matrix of Sect.~\ref{sec:thcovmat}) will be
 left for future studies.

The NNPDF4.0MHOU NNLO PDF set is made publicly available via the
{\sc\small LHAPDF6} interface, 
\begin{center}
  {\bf \url{http://lhapdf.hepforge.org/}~} .
\end{center}
It is  delivered as a set of $N_{\rm rep}=100$ Monte Carlo replicas,
and it is 
denoted as
\begin{flushleft}
  \tt NNPDF40\_nnlo\_as\_01180\_mhou \\
\end{flushleft}
It should be considered a more accurate version of the
published NNPDF4.0 NNLO PDF set~\cite{NNPDF:2021njg}.

This PDF set is also made available via the NNPDF collaboration website
\begin{center}
  {\bf \url{https://nnpdf.mi.infn.it/nnpdf4-0-mhou/}~} ,
\end{center}
where we also make available the other PDF sets presented in
Section~\ref{sec:results}. These include the NLO PDFs with MHOUs based
on the same dataset used at NNLO
\begin{flushleft}
  \tt NNPDF40\_nlo\_as\_01180\_mhou\_nocuts \\
\end{flushleft}
and the corresponding baseline NLO and NNLO sets without MHOUs
\begin{flushleft}
  \tt NNPDF40\_nlo\_as\_01180\_qcd\_nocuts \\
  \tt NNPDF40\_nnlo\_as\_01180\_qcd .\\
\end{flushleft}
The latter NNLO set is equivalent to but differs
from the published NNPDF4.0 NNLO because
of the adoption of a new theory pipeline, and was already presented in
Ref.~\cite{NNPDF:2024djq}  (see in particular Appendix~A of that reference).

The availability of PDFs that include MHOUs in their uncertainty is a
step forward in achieving high-accuracy PDFs that can be used for
precision phenomenology at the percent level. We will soon extend the
results presented here to  approximate
N$^3$LO~\cite{NNPDF:2024nan} (for which approximate results are already
available~\cite{McGowan:2022nag}). This will allow us to discuss the
convergence of the perturbative expansion both of PDFs and
perturbative observables, and the effect of the inclusion of MHOUs upon
it.
Further detailed studies of the phenomenological implications of the
NNPDF4.0 PDFs, including MHOUs and the approximate N$^3$LO PDFs, are
left for future studies~\cite{nnpdfpheno}.

\subsection*{Acknowledgments}

R.~D.~B, L.~D.~D., and R.~S. are supported by the U.K.
Science and Technology Facility Council (STFC) grant ST/T000600/1.
F.~H. is supported by the Academy of Finland
project 358090 and is funded as a part of the Center
of Excellence in Quark Matter of the Academy of Finland, project 346326.
E.~R.~N. is supported by the Italian
Ministry of University and Research (MUR) through the “Rita Levi-Montalcini”
Program. M. U. and Z. K. are supported by the European Research Council under
the European Union's Horizon 2020 research and innovation Programme (grant
agreement n.950246), and partially  by the STFC consolidated grant
ST/T000694/1 and ST/X000664/1.
J.~R. is partially supported by NWO, the Dutch Research Council.
C.~S. is supported by the German Research Foundation (DFG) under reference
number DE 623/6-2.

\appendix
\section{Expansion coefficients}
\label{app:explicit}

We give here explicit expressions for the perturbative expansion
coefficients that
are needed in order to perform scale variation according to the
prescriptions discussed in Section~\ref{sec:theory}. Even though in
this paper we only present results up to NNLO, expressions
up to N$^3$LO are given for future reference.

\paragraph{Running of $a_s$.}
The perturbative solution of \cref{eq:betafnc} is
\begin{align}
    a_s(\lambda \mu^2) &= a_s(\mu^2) - \left(a_s(\mu^2)\right)^2 \beta_0 \ln\lambda + \left(a_s(\mu^2)\right)^3 \left(\left(\beta_0\right)^2 \ln^2\lambda - \beta_1 \ln\lambda \right) \nonumber \\
     &\hspace{20pt} -\left(a_s(\mu^2)\right)^4 \left(\left(\beta_0\right)^3 \ln^3\lambda - \frac 5 2 \beta_0 \beta_1 \ln^2\lambda  + \beta_2 \ln\lambda \right) \nonumber \\
     &\hspace{20pt} + O\left(\left(a_s(\mu^2)\right)^{5}\right). \label{eq:asseries}
\end{align}

\paragraph{PDF evolution.}
The perturbative solution of \cref{eq:DGLAP} is
\begin{align}
    E(\lambda \mu^2 \leftarrow \mu^2) &= \mathbbm 1 - a_s(\mu^2) \gamma_0 \ln\lambda  + \left(a_s(\mu^2)\right)^2 \left[ \frac 1 2 \gamma_0 \left(\beta_0 + \gamma_0\right) \ln^2\lambda - \gamma_1 \ln\lambda \right] \nonumber\\
    &\hspace{20pt} - \left(a_s(\mu^2)\right)^3 \left[ \frac 1 6 \gamma_0 \left(2\left(\beta_0\right)^2 + 3\beta_0\gamma_0 + \left(\gamma_0\right)^2\right) \ln^3\lambda \right. \nonumber\\
    &\hspace{90pt} - \left. \frac 1 6 \left(\beta_1\gamma_0 + 2\beta_0\gamma_1 + \gamma_0\gamma_1 + \gamma_1\gamma_0\right) \ln^2\lambda + \gamma_2 \ln\lambda \right] \nonumber \\
    &\hspace{20pt} + {\cal O}\left(\left(a_s(\mu^2)\right)^{4}\right) \,.\label{eq:ekoseries}
\end{align}

\paragraph{Scale variation of cross-sections and anomalous dimensions.}
The expression of the scale-varied coefficients  $\bar C_j(\rho)$
Eq.~(\ref{eq:Cbarren}) in terms of the expansion coefficients  $ C_j$
Eq.~(\ref{eq:cfexp}) is
\begin{align}
\label{eq:ren_coeff}
    \bar C_0(\rho) &= C_0\,,\\
    \bar C_1(\rho) &= C_1 + m C_0 \beta_0\ln\rho\,, \\
    \bar C_2(\rho) &= C_2 + \frac{m (m+1)}{2}  C_0 \left(\beta_0\right)^2 \ln^2\rho + \left((m+1)C_1 \beta_0 + m C_0 \beta_1 \right)\ln\rho\,, \\
    \bar C_3(\rho) &= C_3 + \frac{m (m+1)(m+2)}{6} C_0 \left(\beta_0\right)^3\ln^3\rho \nonumber\\
     &\hspace{20pt} + \left( \frac{(m+1)(m+2)}{2} C_1 \left(\beta_0\right)^2  + \frac{m(2m+3)}{2} C_0 \beta_0 \beta_1 \right)\ln^2\rho \nonumber\\
     &\hspace{20pt} + \left((m+2)C_2 \beta_0 + (m+1)C_1 \beta_1 + m C_0 \beta_2\right)\ln\rho \,.
\end{align}
The expression for of the scale-varied coefficients  $\bar \gamma_j(\rho)$
Eq.~(\ref{eq:gammabarren}) in terms of the expansion coefficients  $ \gamma_j$
Eq.~(\ref{eq:DGLAPexp}) of course is the same, with $m=1$ and $C\rightarrow\gamma$.

\paragraph{Scale variation of PDFs.}
The expression of the coefficients  $K_j(\rho)$ \cref{eq:ks} in terms
of the expansion coefficients  $ \gamma_j$
Eq.~(\ref{eq:DGLAPexp}) can be obtained by setting $\lambda=1/\rho$ in
Eq.~(\ref{eq:ekoseries}). They are given by
\begin{align}
\label{eq:fact_kernels}
K_0(\rho) &= \mathbbm 1\,,\\
K_1(\rho) &= \gamma_0 \ln\rho\,,\\
K_2(\rho) &= \frac 1 2 \gamma_0 \left(\beta_0 + \gamma_0 \right) \ln^2\rho + \gamma_1 \ln\rho\,,\\
K_3(\rho) &= \frac 1 6 \gamma_0 \left(2\left(\beta_0\right)^2 + 3\beta_0\gamma_0 + \left(\gamma_0\right)^2\right) \ln^3\rho  \nonumber\\
    &\hspace{20pt} + \frac 1 2 \left(\beta_1\gamma_0 + 2\beta_0\gamma_1 + \gamma_0\gamma_1 + \gamma_1\gamma_0\right) \ln^2\rho + \gamma_2 \ln\rho\,.
\end{align}

\paragraph{Factorization scale variation in coefficient functions.}
Substituting Eq.~(\ref{eq:schemebtoc}) in Eq.~(\ref{eq:simple}) the
factorized expression for the physical observable after factorization
scale variation  is
\begin{align}
\label{eq:factc}
F(Q^2) &= C(Q^2)\bar f(Q^2,\rho_f) \nonumber\\
&= C(Q^2)K(a_s(\rho_f Q^2),\rho_f) E(\rho_f Q^2 \leftarrow \mu^2_0) f(\mu_0^2)\\
&=\bar{\bar C}(Q^2,\rho_f) f(\rho_f Q^2)\left[1+O(a_s)\right],
\end{align}
where we defined
\begin{equation}
\label{eq:factccf}
 \bar{\bar C}(Q^2,\rho_f)= C(Q^2)K'(a_s(Q^2),\rho_f)=\a_s^m(Q^2) \sum_{j=0}^k
  \left(a_s(Q^2)\right)^{j} \bar{\bar C}_j(\rho_f),
\end{equation}
and $K'$ is in turn found by  re-expressing $K(\rho_f Q^2,\rho_f)$ as
a series in $a_s(Q^2)$, namely letting
\begin{equation}
K'(a_s(Q^2),\rho_f) = \sum_{j=0}^{k} \left(a_s(Q^2)\right)^{j}K'_j(\rho_f)\label{eq:ks2}
\end{equation}
with the requirement
\begin{equation}
    K(a_s(\rho_f Q^2),\rho_f) = K'(a_s(Q^2),\rho_f)\left[1+O(a_s)\right].
\end{equation}
We get 
\begin{align}
    \label{eq:fact_kernelsprime}
    K'_0(\rho) &= \mathbbm 1\,,\\
    K'_1(\rho) &= \gamma_0 \ln\rho,\\
    K'_2(\rho) &= \frac 1 2 \gamma_0 \left(-\beta_0 + \gamma_0  \right) \ln^2\rho + \gamma_1 \ln\rho\,,\\
    K'_3(\rho) &= \frac 1 6 \gamma_0 \left(2\left(\beta_0\right)^2 - 3\beta_0\gamma_0 + \left(\gamma_0\right)^2\right) \ln^3\rho  \nonumber\\
    &\hspace{20pt} + \frac 1 2 \left(-\beta_1\gamma_0 - 2\beta_0\gamma_1 + \gamma_0\gamma_1 + \gamma_1\gamma_0\right) \ln^2\rho + \gamma_2 \ln\rho\,,
\end{align}
which, substituted in Eq.~(\ref{eq:factccf}) leads to
\begin{align}
    \bar{\bar C}_{0}(\rho) &= C_0\,,\\
    \bar{\bar C}_{1}(\rho) &= C_1 + \gamma_0 C_0 \ln\rho\,,\\
    \bar{\bar C}_{2}(\rho) &= C_2 + \frac{1}{2}\gamma_0(- \beta_0 + \gamma_0)C_0\ln^{2}\rho + \left(\gamma_0C_1+ \gamma_1C_0\right)\ln\rho\,,\\
    \bar{\bar C}_{3}(\rho) &= C_3 + \frac 1 6 \gamma_0 \left(2\left(\beta_0\right)^2 - 3\beta_0\gamma_0 + \left(\gamma_0\right)^2\right) C_0 \ln^3\rho  \nonumber\\
    &\hspace{20pt} + \frac 1 2 \left(-\beta_1\gamma_0 - 2\beta_0\gamma_1 + \gamma_0\gamma_1 + \gamma_1\gamma_0\right) C_0\ln^2\rho \nonumber\\
    &\hspace{20pt} + \frac 1 2 \gamma_0 \left(-\beta_0 + \gamma_0 \right) C_1 \ln^2\rho \nonumber\\
    &\hspace{20pt} + \left(\gamma_0 C_2 + \gamma_1 C_1 +
    \gamma_2 C_0 \right) \ln\rho\,.
\end{align}

\bibliography{nnpdf40mhou}
\end{document}